\documentclass[fleqn,usenatbib]{mnras}

\usepackage{newtxtext,newtxmath}
\usepackage[T1]{fontenc}

\usepackage{amsmath}
\usepackage{stfloats}
\usepackage{graphicx}	

\newcommand{\swift}{\textit{Swift}}

\title[CLEs are TDEs]{Coronal Line Emitters are Tidal Disruption Events in Gas-Rich Environments}

\author[Hinkle et al.]
{\href{http://orcid.org/0000-0001-9668-2920}{Jason T. Hinkle}$^{1}$\thanks{jhinkle6@hawaii.edu}\thanks{FINESST FI},
\href{https://orcid.org/0000-0003-4631-1149}{Benjamin J. Shappee}$^{1}$, and
\href{http://orcid.org/0000-0001-9206-3460}{Thomas W.-S. Holoien}$^{2}$\thanks{NHFP Einstein Fellow} \\
$^{1}$Institute for Astronomy, University of Hawai`i, 2680 Woodlawn Dr., Honolulu, HI 96822, USA\\
$^{2}$The Observatories of the Carnegie Institution for Science, 813 Santa Barbara St., Pasadena, CA 91101, USA \\\
}

\pubyear{2023}

\begin{document}
\label{firstpage}
\pagerange{\pageref{firstpage}--\pageref{lastpage}}
\maketitle

\begin{abstract}
Some galaxies show little to no sign of active galactic nucleus (AGN) activity, yet exhibit strong coronal emission lines (CLs) relative to common narrow emission lines. Many of these CLs have ionization potentials of $\geq 100$ eV, thus requiring strong extreme UV and/or soft X-ray flux. It has long been thought that such events are powered by tidal disruption events (TDEs), but owing to a lack of detailed multi-wavelength follow-up, such a connection has not been firmly made. Here we compare coronal line emitters (CLEs) and TDEs in terms of their host-galaxy and transient properties. We find that the mid-infrared (MIR) colors of CLE hosts in quiescence are similar to TDE hosts. Additionally, many CLEs show evidence of a large dust reprocessing echo in their mid-infrared colors, a sign of significant dust in the nucleus. The stellar masses and star formation rates of the CLE hosts are largely consistent with TDE hosts, with many CLEs residing within the green valley. The blackbody properties of CLEs and TDEs are similar, with some CLEs showing hot (T $\geq 40,000$ K) blackbody temperatures. Finally, the location of CLEs on the peak-luminosity/decline-rate parameter space is much closer to TDEs than many other major classes of nuclear transients. Combined, these provide strong evidence to confirm the previous claims that CLEs are indeed TDEs in gas-rich environments. We additionally propose a stricter threshold of CL flux $\geq 1/3$ $\times$ [\ion{O}{iii}] flux to better exclude AGNs from the sample of CLEs.
\end{abstract}

\begin{keywords}
accretion, accretion discs --- ISM: dust --- galaxies: active --- galaxies: emission lines --- galaxies: nuclei
\end{keywords}

\section{Introduction}

Supermassive black holes (SMBHs) reside in the centers of nearly all massive galaxies \citep[e.g.][]{kormendy95, magorrian98, ho08, kormendy13}. While direct detections of quiescent SMBHs are difficult to obtain, their presence can be illuminated when a star passes within the tidal radius of an SMBH and is torn apart, powering a luminous accretion flare \citep[e.g.,][]{rees88, phinney89, evans89, ulmer99}. Such events are known as tidal disruption events \citep[TDEs; e.g., ][]{vanvelzen11, holoien14b, gezari21}.

Many optically-selected TDEs seem to prefer host galaxies where a recent burst of star formation has occurred \citep[e.g.,][]{stone16b, french16}. These include quiescent Balmer strong and post-starburst galaxies, both likely post-merger systems \citep{french21}. The enhanced TDE rate is likely due to the high stellar densities in the cores of these post-merger galaxies \citep{french20}. However, some studies suggest that the TDE rate can be even higher for systems with higher current star-formation rates, especially luminous infrared galaxies \citep{tadhunter17, mattila18, kool20}, where high nuclear stellar densities may also be present due to the ongoing or very recent merger funneling gas to the nucleus.

TDEs exhibit a wide range of emission across the electromagnetic spectrum, but nevertheless have clear defining features as a class. The majority of TDEs display luminous UV/optical emission which is well-fit by a blackbody with $T \sim 15,000-50,000$ K \citep[e.g.,][]{gezari12b, holoien14b, holoien16a}. Many TDEs also exhibit X-ray emission \citep[e.g.,][]{holoien16a, wevers19, hinkle21a} well-modeled by a hotter blackbody, with kT $\sim30-60$ eV \citep{auchettl17}. TDE light curves often exhibit a smooth rise and monotonic decline \cite[e.g.,][]{auchettl18, holoien20, hinkle21a, vanvelzen21}, generally without stochastic variability. Additionally, there appears to be a correlation between the peak luminosity of a TDE and its decline rate after peak \citep{hinkle20a, hammerstein23}.

One class of transients that has been linked to TDEs are the so-called coronal line emitters \citep[CLEs; ][]{komossa08, wang11, wang12, vanvelzen21b}. Observationally, these events are characterized by strong high ionization coronal lines (CLs) such as [\ion{Fe}{x}] $\lambda$6376, [\ion{Fe}{xi}] $\lambda$7894, [\ion{Fe}{xiv}] $\lambda$5304, and [\ion{S}{xii}] $\lambda$7612, with ionization potentials of up to several hundred eV. In particular CLEs are characterized by CLs with strong emission relative to common AGN lines such as [\ion{O}{iii}] $\lambda$5007 \citep[e.g.,][]{wang12}. While AGNs are known to sometimes power strong high ionization CL emission \citep[e.g.,][]{negus21, cerqueiracampos21, prieto22}, the spectral shapes and emission line ratios of these CLEs are generally not consistent with strong AGNs.

The most likely explanation for these events are TDEs occurring in gas-rich environments, where the strong extreme UV and soft X-ray flux from TDEs can photoionize gas in the host galaxy. To date, the proposed connection of CLEs to TDEs has been based on the transient nature of the coronal line emission \citep{komossa09, wang12, yang13}, broad \ion{He}{ii} emission consistent TDE spectral features \citep{wang11, wang12, yang13}, a TDE-like decline slope for a small number of cases \citep{palaversa16}, and similar rate estimates \citep{wang12}. The detection of increasing [\ion{O}{iii}] flux on $\sim 5 - 7$ year timescales in some CLEs strongly supports the existence of significant gas in the nuclei of the CLE host galaxies \citep{yang13}.

In this work, we strengthen the proposed link between CLEs and TDEs by leveraging nearly a decade of fruitful searches for TDEs. In Section \ref{sample} we discuss our CLE sample selection in addition to our comparison sample of TDEs. We compare the host galaxy properties of CLEs to those of TDEs in Section \ref{hosts} and compare the spectral energy distributions of the transients in Section \ref{seds}. Then, in Section \ref{sec:cls} we discuss the selection methods for CLEs. Finally, in Section \ref{conc}, we provide a discussion and our conclusions. Throughout the paper we assume a cosmology of $H_0$ = 69.6 km s$^{-1}$ Mpc$^{-1}$, $\Omega_{M} = 0.29$, and $\Omega_{\Lambda} = 0.71$ \citep{wright06, bennett14}.

\begin{figure*}
\centering
 \includegraphics[width=0.98\textwidth]{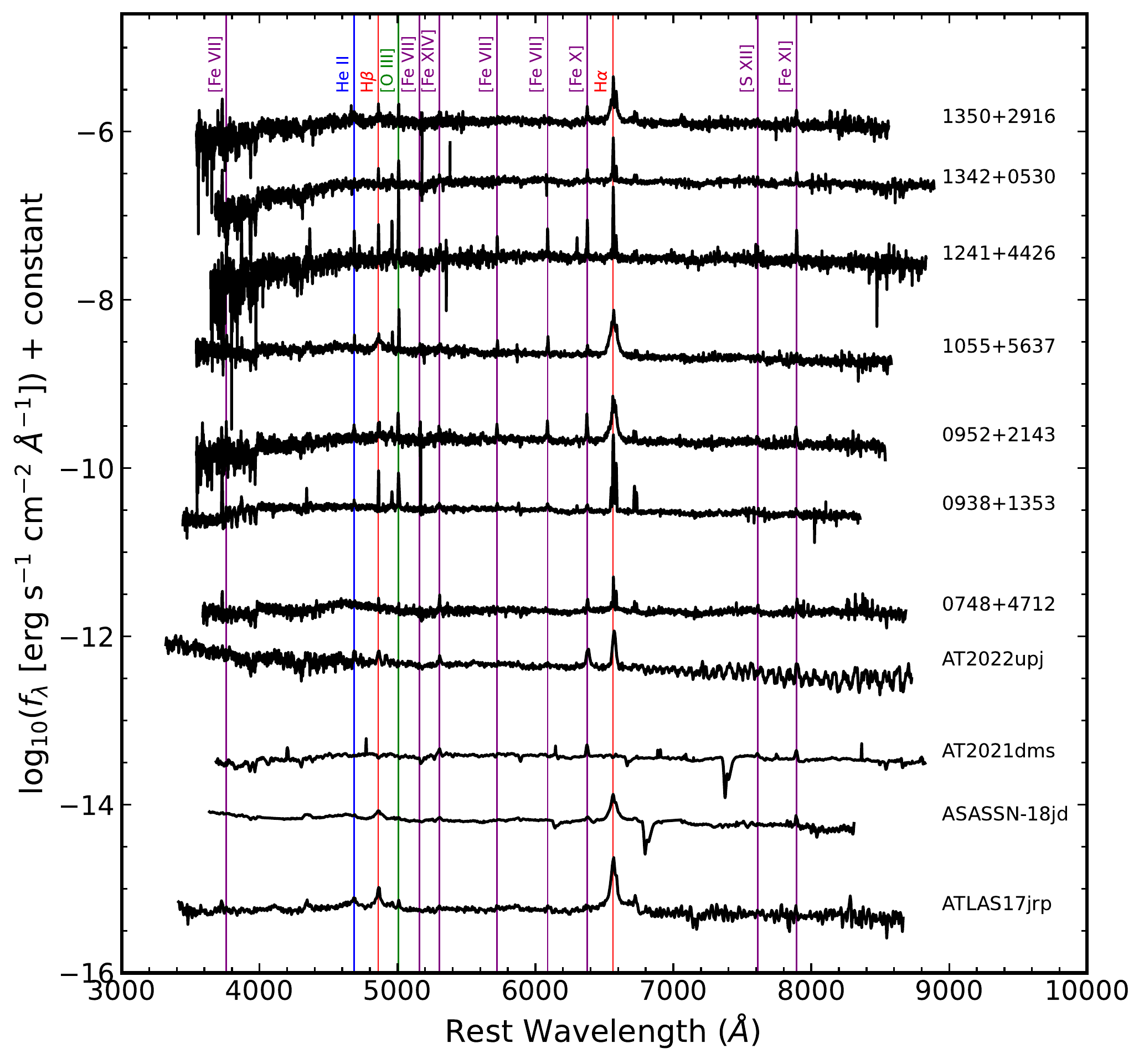}\hfill
 \caption{Optical spectra of our sample of CLEs. The vertical lines mark spectral features common in AGNs and TDEs, with hydrogen lines in red, helium lines in blue, and oxygen lines in green. Selected coronal lines of iron and sulfur are shown in purple, which can be relatively weak in sources without significant [\ion{O}{iii}] emission. The source names are given at the right of each spectrum.}
 \label{fig:spectra}
\end{figure*}

\section{Sample Selection} \label{sample}

In this section, we discuss our sample selection of CLEs from spectroscopic and photometric surveys. We also introduce our comparison sample of TDEs.

\subsection{Coronal Line Emitters}

Our sample of CLEs comes from two distinct selection methods which we will call photometrically-selected and spectroscopically-selected CLEs throughout this work.

The photometrically-selected CLEs are sources that were initially discovered based on a large-scale rise in the continuum flux and then classified through spectroscopic follow-up. In our search for such objects, we include recent discoveries of singular nuclear flares from all-sky surveys for which classification or follow-up spectra reveal coronal line emission that is strong relative to narrow AGN lines. We find 4 events, which were initially discovered based on large-scale continuum flux changes and were selected in a similar manner to other optically-selected TDEs, resulting in initial spectroscopic classifications as TDE candidates. In this manuscript, we will denote the CLEs discovered in sky surveys as photometrically-selected CLEs. These are ATLAS17jrp \citep{onori22, wang22b}, ASASSN-18jd \citep{neustadt20}, AT2021dms, and AT2022upj \citep{newsome22, fulton22}.

The spectroscopically-selected CLEs are events that were initially discovered based on emission line properties in a spectrum, often without accompanying photometry. In particular, this class includes the spectroscopically-selected CLEs from SDSS \citep[e.g.,][]{komossa08, wang11, wang12}. These were selected by systematically searching the archive of SDSS spectra for sources with coronal lines strong relative to typical narrow AGN emission lines, such as [\ion{O}{iii}] $\lambda$5007, and removing sources classified as AGNs through typical line diagnostic diagrams, yielding 7 CLEs. We will refer to these CLEs as spectroscopically-selected in the remainder of this manuscript. Throughout the manuscript, we will shorten the names of the SDSS sample of CLEs to improve readability (e.g. SDSSJ074820.66+471214.6 becomes 0748+4712). The full sample of 11 events is listed in Table \ref{tab:sample}. 

The optical spectra for our sample of CLEs are shown in Figure \ref{fig:spectra}. For the SDSS sample, we use the SDSS spectra \citep{aguado19}, the same as shown in \citet{wang12}. For ATLAS17jrp, we show a publicly available ePESSTO \citep{smartt15, barbarino19} spectrum taken with the ESO Faint Object Spectrograph and Camera (EFOSC2) from the European Southern Observatory archive. For ASASSN-18jd we show a Southern African Large Telescope (SALT) spectrum from \citet{neustadt20}. We obtained a spectrum of AT2021dms using the Low Dispersion Survey Spectrograph (LDSS-3) on the Magellan Clay telescope, taken on MJD 59439.3. This new spectrum was reduced using standard IRAF \citep{tody86, tody93} procedures, such as bias subtraction, flat-fielding, wavelength calibration with arc lamps and flux calibration with spectrophotometric standard stars. Finally, for AT2022upj, we use a classification spectrum on the Transient Name Server obtained with the FLOYDS-S spectrograph on Faulkes Telescope South \citep{newsome22}. The CLE nature of these transients is confirmed by spectroscopic similarity to the spectroscopically-selected CLEs.

\begin{table*}
\centering
 \caption{Sample of Coronal Line Emitters}
 \label{tab:sample}
 \begin{tabular}{cccccc}
  \hline
  Object & TNS ID & $z$ & Right Ascension & Declination & References \\
  \hline
  SDSSJ074820.66+471214.6 & --- & 0.0615 & 07:48:20.674 & $+$47:12:14.30 & \citet{wang11, wang12} \\
  SDSS J093801.64+135317.0 & --- & 0.1006 & 09:38:01.636 & $+$13:53:17.07 & \citet{wang12} \\
  SDSS J095209.56+214313.3 & --- & 0.07949 & 09:52:09.563 & $+$21:43:13.31 & \citet{komossa08, wang12} \\
  SDSS J105526.43+563713.3 & --- & 0.07396 & 10:55:26.411 & $+$56:37:13.04 & \citet{wang12} \\
  SDSS J124134.26+442639.2 & --- & 0.0419 & 12:41:34.253 & $+$44:26:39.25 & \citet{wang12} \\
  SDSS J134244.42+053056.1 & --- & 0.03655 & 13:42:44.416 & $+$05:30:56.13 & \citet{wang12} \\
  SDSS J135001.49+291609.7 & --- & 0.07768 & 13:50:01.500 & $+$29:16:09.70 & \citet{wang12} \\
  ATLAS17jrp & AT2017gge & 0.066 & 16:20:35.004 & $+$24:07:26.57 & \citet{onori22, wang22b} \\
  ASASSN-18jd & AT2018bcb & 0.1192 & 22:43:42.871  & $-$16:59:08.49 & \citet{neustadt20} \\
  ZTF20acsacog & AT2021dms & 0.031102 & 03:21:24.069	& $-$11:08:45.71	 & \citet{forster21} \\
  ZTF22abegjtx & AT2022upj & 0.054 & 00:23:56.846 & $-$14:25:23.22	 & \citet{newsome22, fulton22}\\
  \hline
 \end{tabular}\\
\begin{flushleft} The 11 CLEs analyzed in this manuscript. Those with full SDSS names are spectroscopically-selected and those with survey names are photometrically-selected. The TNS ID is the identification given for objects reported on the Transient Name Server, and is only given for the photometrically-selected CLEs. The references are the discovery paper and key follow-up efforts where available or the transient announcement/classification as needed.
\end{flushleft}
\end{table*}

\subsection{Tidal Disruption Event Comparison Sample}

Within the past several years the sample of well-studied TDEs has grown considerably. In particular, the number of TDEs discovered prior to their peak and with significant multi-wavelength follow-up has increased dramatically. In this manuscript we compare our CLEs to the sample of \citet{hinkle21b}, as these sources each have significant multi-wavelength data and have been reduced in a uniform manner. That work compiled observations of 27 well-studied TDEs \citep{holoien14b, holoien16a, holoien16b, hung17, wyrzykowski17, blagorodnova17, blagorodnova18, holoien19b, holoien19c, leloudas19, nicholl20, wevers19, vanvelzen21, hinkle21a}. 

\section{Comparison of Host-Galaxy Properties} \label{hosts}

We first compare the host galaxies of our sample of CLEs and TDEs. In Section \ref{sec:wise} we examine archival WISE photometry and in Section \ref{sec:stellar} we examine key host properties such as stellar mass and star-formation rate (SFR).

\subsection{WISE}
\label{sec:wise}

Mid-infrared (MIR) photometry has long been used to distinguish AGNs from normal galaxies based on color cuts \citep[e.g.,][]{stern04, assef10, assef13}. The all-sky Wide-field Infrared Survey Explorer \citep[WISE; ][]{wright10} has been especially useful in selecting AGNs, as it mapped the entire sky in four mid-infrared (MIR) bands: $W1$ (3.4 $\mu m$), $W2$ (4.6 $\mu m$), $W3$ (12 $\mu m$) and $W4$ (22 $\mu m$). As compared to inactive galaxies, AGNs show redder WISE colors due to the strength of the non-thermal power-law component of the AGN SED as compared to the stellar population of the galaxy. 

\begin{figure}
\centering
 \includegraphics[width=0.48\textwidth]{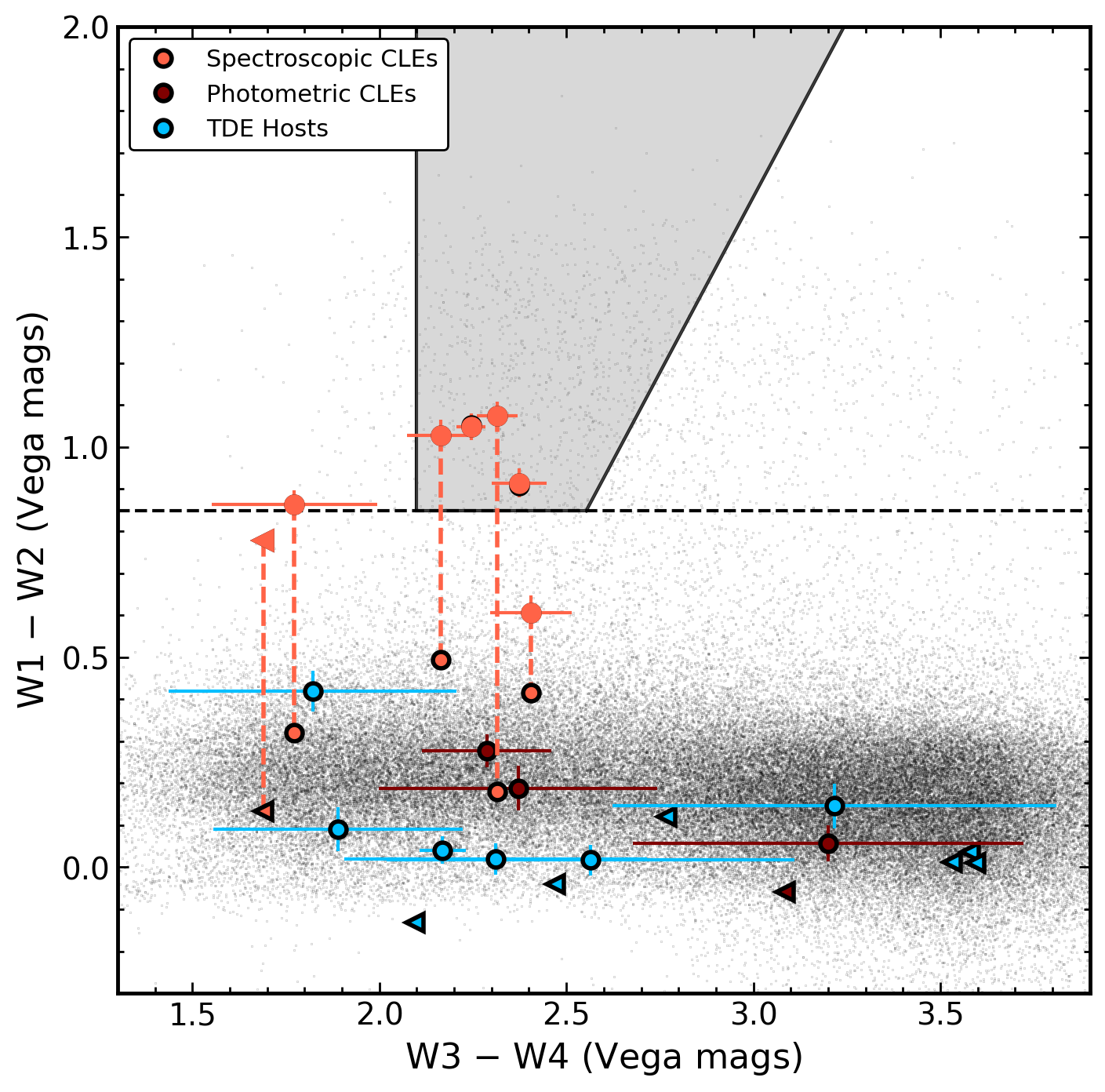}\hfill
 \caption{WISE color-color diagram used to discriminate strong AGNs from star-forming galaxies. The gray box is the AGN region determined by \citet{assef13}. The orange points are the spectroscopically-selected CLEs, the dark red points are the photometrically-selected CLEs, and the blue points are the comparison sample of TDEs. The orange points without black borders use the AllWISE catalog photometry and are connected by a dashed line to points with $W1 - W2$ colors taken from the most recent epoch of NEOWISE photometry. The background points are a sample of HyperLeda \citep{makarov14} galaxies shown to illustrate the distribution of normal galaxies.}
 \label{fig:wise_colors}
\end{figure}

In Figure \ref{fig:wise_colors} we show CLE and TDE hosts in a common WISE color-color selection diagram \citep{assef13}. In general, AGNs lie above a $W1 - W2$ color of $\sim$0.8 Vega mag. Here we see that, like the other TDEs, the photometrically-selected CLEs have WISE colors typical of non-active galaxies. Simply using the AllWISE catalog \citep{wright10} photometry (data points without black borders in Fig. \ref{fig:wise_colors}), it appears that nearly all the spectroscopically-selected CLEs would be selected as AGNs. However, it is likely that many hosted transient events \citep[e.g.,][]{yang13, palaversa16} temporally overlapping with the AllWISE mapping.

This raises the possibility that these AGN-like WISE colors were the result of nuclear dust heated by a transient event \citep{lu16, vanvelzen16b, wang18, jiang21a, hinkle22b}. Fortunately, NEOWISE \citep{mainzer14}, the successor mission to WISE, was activated in October 2013. This allows for continued MIR mapping of the sky, albeit at a low cadence and only in the $W1$ and $W2$ filters. If we use the most recent NEOWISE epoch to compute the $W1 - W2$ color for these spectroscopically-selected CLEs instead of AllWISE, we find the majority of the spectroscopically-selected CLEs appear to be normal galaxies in the WISE color-color space. Nevertheless, two sources, 0938+1353 and 1055+5637, do not show significant changes in their $W1 - W2$ color. 

\begin{figure}
\centering
 \includegraphics[width=0.48\textwidth]{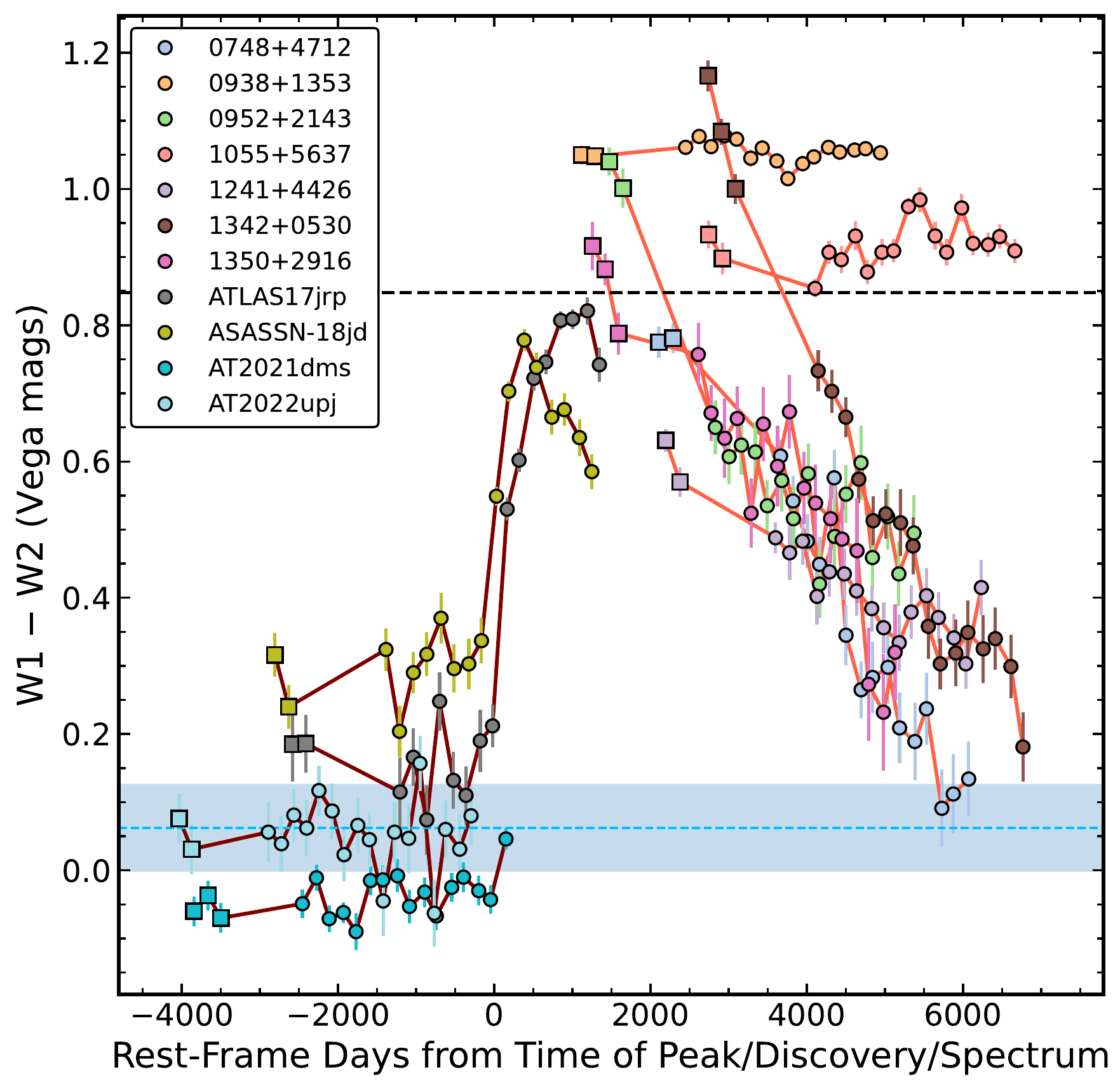}\hfill
 \caption{WISE $W1 - W2$ color curves for our sample of CLEs. AllWISE photometry is shown as squares and NEOWISE photometry is shown as circles. The orange lines are spectroscopically-selected CLEs with time relative to the SDSS spectrum and the dark red lines indicate photometrically-selected CLEs with time relative to the peak/discovery time. The individual point colors denote separate events as shown in the legend. The dashed black line is the typical AGN color cut, above which sources likely host an AGN. The dashed blue line and shaded region are the median and 1$\sigma$ range of the WISE $W1 - W2$ colors of the TDE comparison sample.}
 \label{fig:wise_lcs}
\end{figure}

To further test the possibility of transient-heated dust, we made WISE light curves of these events, combining the individual AllWISE epochs with NEOWISE survey photometry. To create our NEOWISE light curves, we stacked the intra-visit exposures to create deeper $W1$ and $W2$ light curves with $\sim$6 month cadence. We show long-term WISE color curves in Figure \ref{fig:wise_lcs}. The WISE color curves for most of the spectroscopically-selected CLEs become bluer over time, consistent with transient heating of nuclear dust. The $W1$ and $W2$ light curves for these sources also show declines. Our interpretation of the declining WISE light curves and $W1 - W2$ color evolution as nuclear dust heated by a TDE agrees with theoretical work on TDE dust echoes \citep[e.g.,][]{lu16}. Additionally, our results are in excellent agreement with previous work on the MIR properties of transient CLEs \citep{dou16}.

In contrast to the typical behavior of the sample, the CLEs 0938+1353 and 1055+5637 show relatively flat WISE colors, with small-scale variability consistent with AGN activity \citep{sheng17}. Like their WISE colors, the  $W1$ and $W2$ light curves for 0938+1353 and 1055+5637 are flat, without any consistent decline. \citet{yang13} also find that both of these sources do not exhibit significant changes in their CL emission over long baselines, likely suggesting persistent AGN activity. We will discuss these sources in more detail in Section \ref{sec:cls}. We will treat 0938+1353 and 1055+5637 differently from the other spectroscopically-selected CLEs as they are likely related to AGN activity rather than a transient event.

The early evolution of the photometrically-selected CLEs all show WISE colors typical of normal galaxies and are consistent with the sample of TDE hosts. ASASSN-18jd is slightly redder than the typical TDE host, but not to a level where strong AGN emission must be present. Some photometrically-selected CLEs show signs of large-scale dust echoes at late times; these are particularly notable for ATLAS17jrp and ASASSN-18jd, whose WISE colors grow significantly redder after their peak emission. These dust echoes indicate significant dust in the nucleus \citep{wang18, jiang21a, hinkle22b}. Given a typical gas-to-dust ratio of $\sim$100 in the interstellar medium \citep[e.g.,][]{hildebrand83, draine84}, this implies significant amounts of gas present in the nuclei of these galaxies. As AT2021dms and AT2022upj are relatively recent events, their dust echoes are likely still rising and future NEOWISE releases will likely reveal dust reprocessing echoes, as suggested in Figure \ref{fig:wise_lcs}.

\subsection{Stellar Mass and Star Formation Rate}
\label{sec:stellar}

\begin{figure}
\centering
 \includegraphics[width=0.48\textwidth]{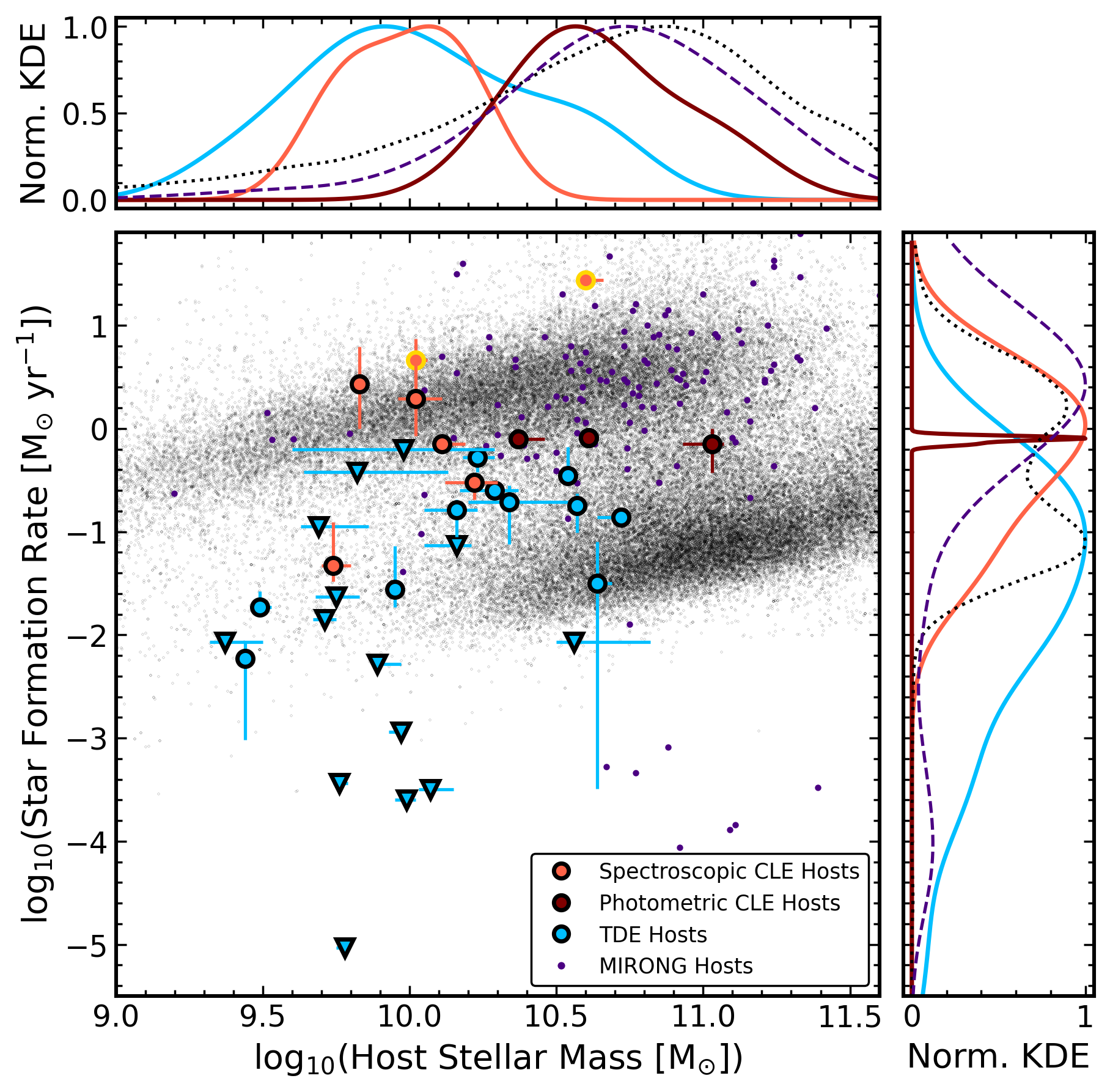}\hfill
 \caption{Host-galaxy star formation rate as compared to stellar mass. The orange points are the spectroscopically-selected CLEs, the dark red points are the photometrically-selected CLEs, and the blue points are the comparison sample of TDEs. Two photometrically-selected CLEs (AT2021dms and AT2022upj) have nearly identical masses and SFRs. Upper-limits on SFRs are shown as downward-facing triangles. The two points with golden borders are 0938+1353 and 1055+5637, which each show evidence for AGN activity. In the two outer panels, we show normalized KDEs of the distributions for the mass and SFRs of the two sub-samples of CLEs and the TDE comparison sample. We have excluded 0938+1353 and 1055+5637 from the spectroscopically-selected CLE KDEs. The background gray points are a sample of SDSS DR8 \citep{eisenstein11} galaxies with properties taken from the MPA-JHU catalog \citep{brinchmann04} and the background indigo points are the sample of MIRONGs from \citet{jiang21b} with corresponding KDEs as the dotted and dashed lines respectively.}
 \label{fig:host_props}
\end{figure}

While the WISE colors provide insight into the possible presence of AGN activity and the heating of nuclear dust, we are also interested in the bulk properties of the galaxies in which these CLEs reside. As such, we next fit archival photometry of the host galaxies with the Fitting and Assessment of Synthetic Templates \citep[\textsc{Fast};][]{kriek09} as we have done for previous transient events \citep[e.g.,][]{holoien16a, hinkle21a, hinkle21b}. When available, we obtained photometry in the UV from the Galaxy Evolution Explorer \citep[GALEX;][]{martin05}, optical photometry from SDSS \citep{aguado19} or Pan-STARRS \citep{chambers16, flewelling16}, and infrared photometry from the Two Micron All-Sky Survey \citep[2MASS;][]{skrutskie06} and AllWISE catalogs.

However, as the spectroscopically-selected CLEs clearly have contamination in their AllWISE data, we have excluded it from their host-galaxy fits. We also consider the survey timescales for the other sources of our galaxy photometry. GALEX mapped the sky from 2003 to 2013, SDSS conducted its imaging survey between 2000 and 2005, and 2MASS took images between 1997 and 2001. Therefore, while there is mild overlap between some GALEX and SDSS images and the times at which the SDSS spectra of our CLEs were taken, the likelihood of significant transient flux in the final photometry is low. This is especially true as the SDSS spectra do not show significant optical continuum excesses that indicate strong transient flux contributions. The GALEX and 2MASS data are most important for constraining SFR and stellar mass from our SED fits respectively, and each has a low probability of transient flux significantly affecting the photometry. We also compared our \textsc{Fast} results with estimates for the spectroscopically-selected CLEs available in the MPA-JHU catalog \citep{brinchmann04} and generally find excellent agreement in mass and good agreement in SFR, although the emission-line based SFR estimates are occasionally higher than our estimates.

In Figure \ref{fig:host_props} we compare the CLE hosts to a sample of TDEs with host-galaxy properties computed in a similar manner by \citet{hinkle21b}. We also compare our CLE hosts to a broader galaxy sample from the MPA-JHU catalog \citep{brinchmann04} and the sample of Mid-infrared Outbursts in Nearby Galaxies (MIRONGs) from \citet{jiang21a}. While the SFR estimates for the SDSS galaxies are computed from spectra, they provide a large sample of uniformly-computed galaxy properties. In addition to the individual source host properties, we show kernel density estimates (KDE) computed using \textsc{scipy.stats.gaussian\_kde} and Scott's Rule to estimate the underlying distribution of these properties for each sub-sample, excluding the sources 0938+1353 and 1055+5637.

In terms of stellar mass, we find that the CLE hosts are generally similar to TDE hosts, but prefer the more massive end of the TDE distribution. Interestingly, of the CLE sample, the photometrically-selected sub-sample resides at the very highest end of the TDE stellar mass distribution, with ASASSN-18jd lying well above the stellar mass of any other TDE or CLE in our samples. The stellar masses of the photometrically-selected CLEs are quite similar to the masses seen for the MIRONG hosts \citep{jiang21a} Conversely, for star-formation rate, the CLE hosts from both selection methods prefer higher star-formation rates as compared to TDE hosts, although they are consistent in many cases with the TDE hosts with the highest SFRs. While some of the CLE hosts have similar SFRs to the MIRONG hosts, the MIRONGs generally lie well within the higher mass end of the blue sequence, at higher SFRs than the typical CLE. Some spectroscopically-selected CLEs have significantly higher SFRs than TDEs, especially those with evidence for AGN activity. The higher SFRs of CLEs is not a large surprise as gas is needed for CL emission to be seen and many TDE hosts are known to be relatively gas-poor \citep[e.g.,][]{french17, french20, jiang21b}. 

Overall, accounting for the need for increased nuclear gas, the host properties of CLEs are similar to TDE hosts in both mass and SFR. Perhaps most interesting when comparing the CLE hosts to the distribution of SDSS galaxies in mass and SFR is that many seem to reside either in the green valley or the lower edge of the blue sequence, much like TDE hosts \citep[e.g..,][]{french20, hammerstein21}. Nevertheless, four of our CLEs clearly reside in star-forming galaxies. This is particularly true for the CLEs that have signs of AGN activity, shown in Fig. \ref{fig:host_props} as points with golden borders.

To evaluate the potential contribution of AGN activity to the observed emission of the CLEs and their host galaxies, we use the Code Investigating GALaxy Emission \citep[CIGALE;][]{burgarella05, boquien19, yang22}. We fit the same data as was used for our FAST fits and employ a similar model with an exponentially-declining star formation history, the \citep{bruzual03} stellar population models, a Salpeter initial mass function \citep{salpeter55}, and a \citet{cardelli89} extinction law with R$_V$ = 3.1. The fits adopt the SKIRTOR AGN model \citep{stalevski12, stalevksi16}. From our CIGALE fits, we find that only two of the CLE hosts in our sample have AGN luminosities comparable to the stellar luminosity. These are 0938+1353, with an AGN-to-stellar luminosity ratio of 0.94 and 1241+4426 with a ratio of 1.19. Interestingly, 1055+5637, one of two sources with no considerable evolution in W1 $-$ W2 color, potentially a sign of AGN activity, has a relatively small luminosity ratio of 0.23.

\begin{figure*}
\centering
 \includegraphics[width=0.98\textwidth]{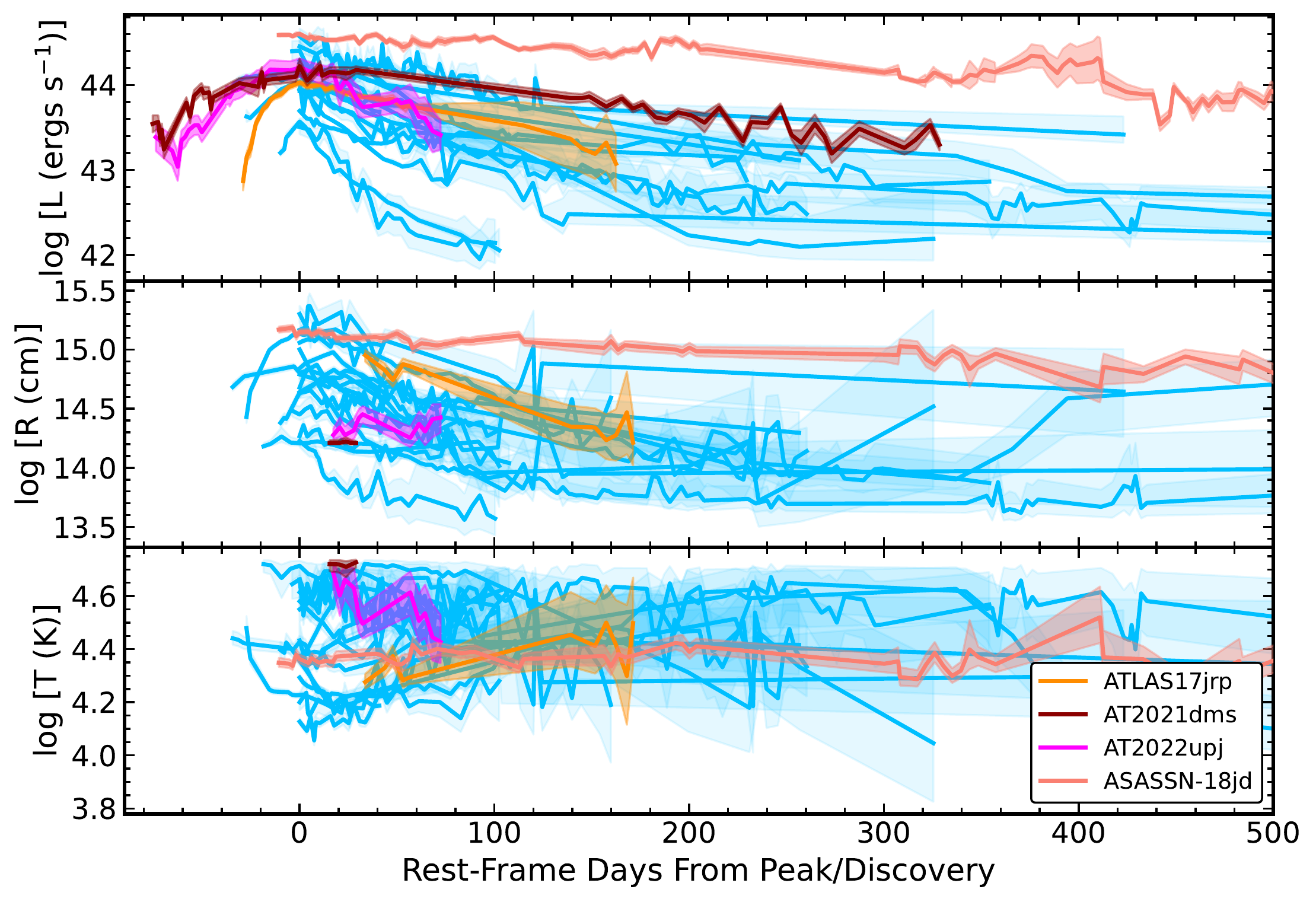}\hfill
 \caption{Temporal evolution of the UV/optical blackbody luminosity (top panel), radius (middle panel), and temperature (bottom panel) for the photometrically-selected CLEs (red shades) and the comparsion sample of TDEs (blue). The solid lines are the median values and the semi-transparent shading corresponds to the 1$\sigma$ uncertainty. Time is in rest-frame days relative to the earliest \swift\ epoch. We have not shown the very-late time blackbody properties for several comparison TDEs.}
 \label{fig:BB_fits}
\end{figure*}

\section{Comparison of SED Properties} \label{seds}

In addition to the properties of the host galaxies, we can compare the properties of the transient events themselves. Unfortunately, this is only doable for the photometrically-selected sub-sample, for which real-time follow-up was conducted and multi-wavelength coverage exists. In this section, we compare the evolution of blackbody parameters and decline rates of these CLEs to well-studied TDEs.

\subsection{Blackbody Parameters}

Each of our photometrically-selected CLEs has follow-up imaging from the \textit{Neil Gehrels Swift Observatory} \citep[\swift;][]{gehrels04} UltraViolet and Optical Telescope \citep[UVOT;][]{roming05}. Many of these epochs use the six typical filters \citep{poole08}: $V$ (5425.3 \AA), $B$ (4349.6 \AA), $U$ (3467.1 \AA), $UVW1$ (2580.8 \AA), $UVM2$ (2246.4 \AA), and $UVW2$ (2054.6 \AA), where the wavelengths quoted here are the pivot wavelengths calculated by the SVO Filter Profile Service \citep{rodrigo12}.

Two sources have had \swift\ UVOT photometry and blackbody fits published in the literature. These are ASASSN-18jd \citep{hinkle21b} and ATLAS17jrp \citep{hinkle22b}, which we adopt in this manuscript. For the other sources, we reduced the UVOT data following the procedures of \citet{hinkle21b}, first summing the individual exposures per epoch in each filter. For AT2021dms and AT2022upj, we used our \textsc{Fast} SED fits and computed synthetic photometry using the corresponding filter response functions to estimate synthetic host fluxes which we then subtracted from our reduced \swift. For AT2021dms (PI: Hinkle) and AT2022upj (PIs: Newsome \& Wang), we computed UVOT magnitudes using a radius that best matched the radius used for the archival host photometry. While we do not include the spectroscopically-selected CLEs here, as their time of peak emission is highly uncertain and each has little UV follow-up data available, the SDSS CLE 0952+2143 was shown to have modest UV emission even at late times \citep{palaversa16}, consistent with a TDE.

We then fit each of these host-subtracted and Galactic extinction-corrected UVOT epochs using a blackbody model, similar to many previous TDEs \citep{holoien14b, holoien16a, vanvelzen21, hinkle21b}. For each epoch of host-subtracted UV photometry, we used Markov Chain Monte Carlo (MCMC) methods and a forward-modeling approach to fit a blackbody model. Figure \ref{fig:BB_fits} shows these fits in comparison to fits of well-observed TDEs from \citet{hinkle21b}. 

Similar to previous TDEs \citep{holoien20, hinkle21a, hinkle21b}, we bolometrically corrected optical survey light curves using the fitted blackbody luminosity. This was done by scaling either the ATLAS $o$ (for ATLAS17jrp, ASASSN-18jd, and AT2021dms) or ZTF $g$ (for AT2022upj) light curves to match the bolometric luminosity estimated from the blackbody fits. In between \swift\ UVOT epochs, the scaling is interpolated. Outside of this range, we assume the correction factor of the nearest \swift\ point, effectively assuming a constant blackbody temperature. This is reasonable for most TDEs \citep[e.g.,][]{hinkle21b, vanvelzen21}, but introduces a possible source of systematic uncertainty. We show these bolometric luminosity curves in the top panel of Fig. \ref{fig:BB_fits}.

Unfortunately, AT2021dms and AT2022upj do not have particularly long \swift\ UVOT coverage. Nevertheless, after bolometrically correcting the optical light curves it is clear that bolometric luminosities of the CLEs and TDEs are generally similar. Of the CLE photometric sample, only ASASSN-18jd is especially luminous, but even so it is consistent with the most luminous TDEs in our comparison sample at peak. It is also well within the luminosity range seen for the recent class of featureless TDEs \citep{hammerstein23}. While the temporal coverage is much shorter on average than the TDEs, the CLE blackbody radii and temperatures are similar to TDEs as well. ASASSN-18jd has a larger radius than most TDEs, but its temperature is fully consistent with the cooler end of the TDE temperature distribution. We additionally note that 2 of the CLEs have blackbody temperatures among the highest seen for any TDE. As the production of CL emission requires significant extreme UV flux, this may not be surprising. However, as other CLEs have relatively modest blackbody temperatures, CL emission cannot rely solely on extremely hot thermal emission.

\subsection{Decline Rates}

\begin{figure}
\centering
 \includegraphics[width=0.48\textwidth]{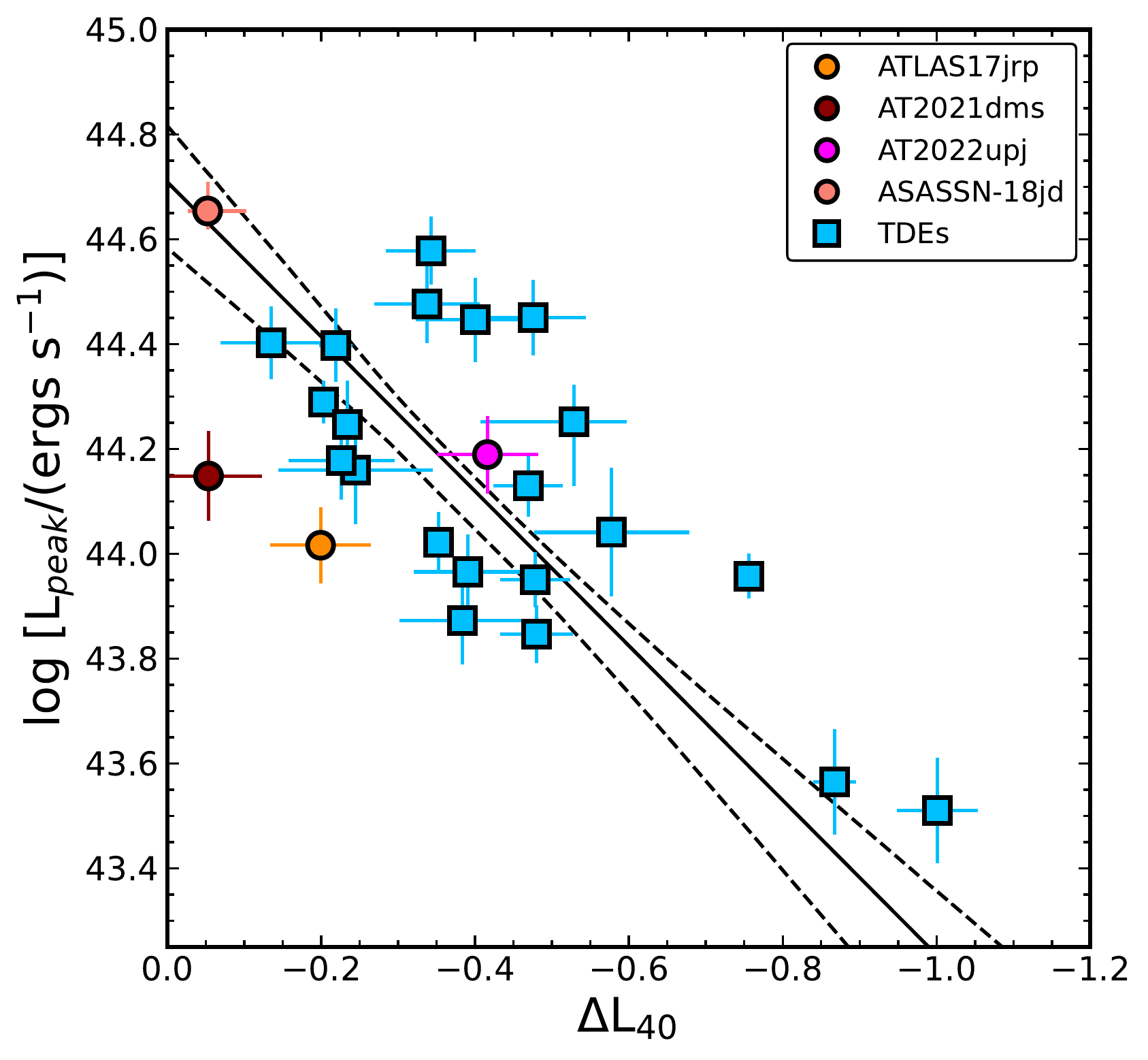}\hfill
 \caption{The peak luminosity versus the decline rate of the photometrically-selected CLEs (red shades) and TDEs (blue). The uncertainties on the CLEs are likely underestimated as for most objects the temperature constraint comes after the time of peak luminosity. The decline rate $\Delta$L$_{40}$ is the difference between the log of the peak luminosity and the log of the luminosity at 40 days after peak. The black solid line is the best-fit line for the TDEs \citep{hinkle21b} and the dashed black lines are the allowed range of uncertainty from the best-fit line.}
 \label{fig:deltaL40}
\end{figure}

It has been noted previously \citep{hinkle21a, hinkle21b, hammerstein23} that there is a relationship between the peak luminosity of a TDE and its bolometric decline rate, namely that more luminous TDEs decline more slowly after peak. As such, we test if the photometrially-selected CLEs in our sample follow a similar trend to the TDEs. Following the procedures of \citet{hinkle20a} we measure the peak bolometric luminosity and decline rate at 40 days after peak. Figure \ref{fig:deltaL40} shows this measurement for our CLEs as compared to the comparison sample of TDEs. For these CLEs, the uncertainties on the peak luminosity and decline rate are likely underestimated as the temperature estimate comes from \swift\ photometry taken after peak emission. We add a 15\% error in quadrature to the peak luminosity values for ATLAS17jrp, AT2021dms, and AT2022upj following the estimates of \citet{hinkle20a} for objects without \swift\ data at peak.

The CLEs straddle the peak-luminosity/decline-rate relationship of TDEs. Previous studies have shown the positions of various classes of supernovae \citep{hinkle20a} and ambiguous nuclear transients \citep[ANTs; ][]{hinkle22a} in this space. Both ANTs and supernovae reside off the TDE relationship, unlike the CLEs in our sample. For ANTs, the median distance is 4.3$\sigma$ and for SNe the median distance is 1.8$\sigma$. When including all the CLEs, the median offset is 1.2$\sigma$, substantially less than both ANTs and SNe. Nevertheless, 2 of our 4 photometrically-selected CLEs lie below the best-fit line, at a level of $2.1\sigma$ for ATLAS17jrp and $2.6\sigma$ for AT2021dms, solely accounting for the intrinsic scatter in the relationship computed by \citet{hinkle21b} and excluding the uncertainties on the points themselves. This may be a natural consequence of the somewhat more massive hosts in which they reside which presumably also host more massive SMBHs for which the fallback timescale is increased. Alternatively, small amounts of dust along the line of sight may decrease the peak luminosity for some CLEs as compared to the nearly dust-free TDE hosts.

\section{Coronal Line Selection} \label{sec:cls}

In \citet{wang12}, one of the key selection criteria for CLEs was at least one CL with a flux of more than 20\% that of the [\ion{O}{iii}] emission line. This selected sources with strong coronal lines, but did seem to allow a fraction of AGNs into the sample, as evidenced by the flat CL emission over a several-year baseline for some sources \citep[e.g.][]{yang13}. The CLEs 0938+1353 and 1055+5637, without fading CLs, are also shown in Fig. \ref{fig:wise_lcs} to have flat and red WISE light curves consistent with AGNs. This suggests that even when attempting to restrict the sample of CLEs to exclude strong AGNs, which can power CLs in their own right, that a population of [\ion{O}{iii}]-weak sources may exist that are not related to transient events, but rather unusual AGNs.

\begin{figure}
\centering
 \includegraphics[width=0.48\textwidth]{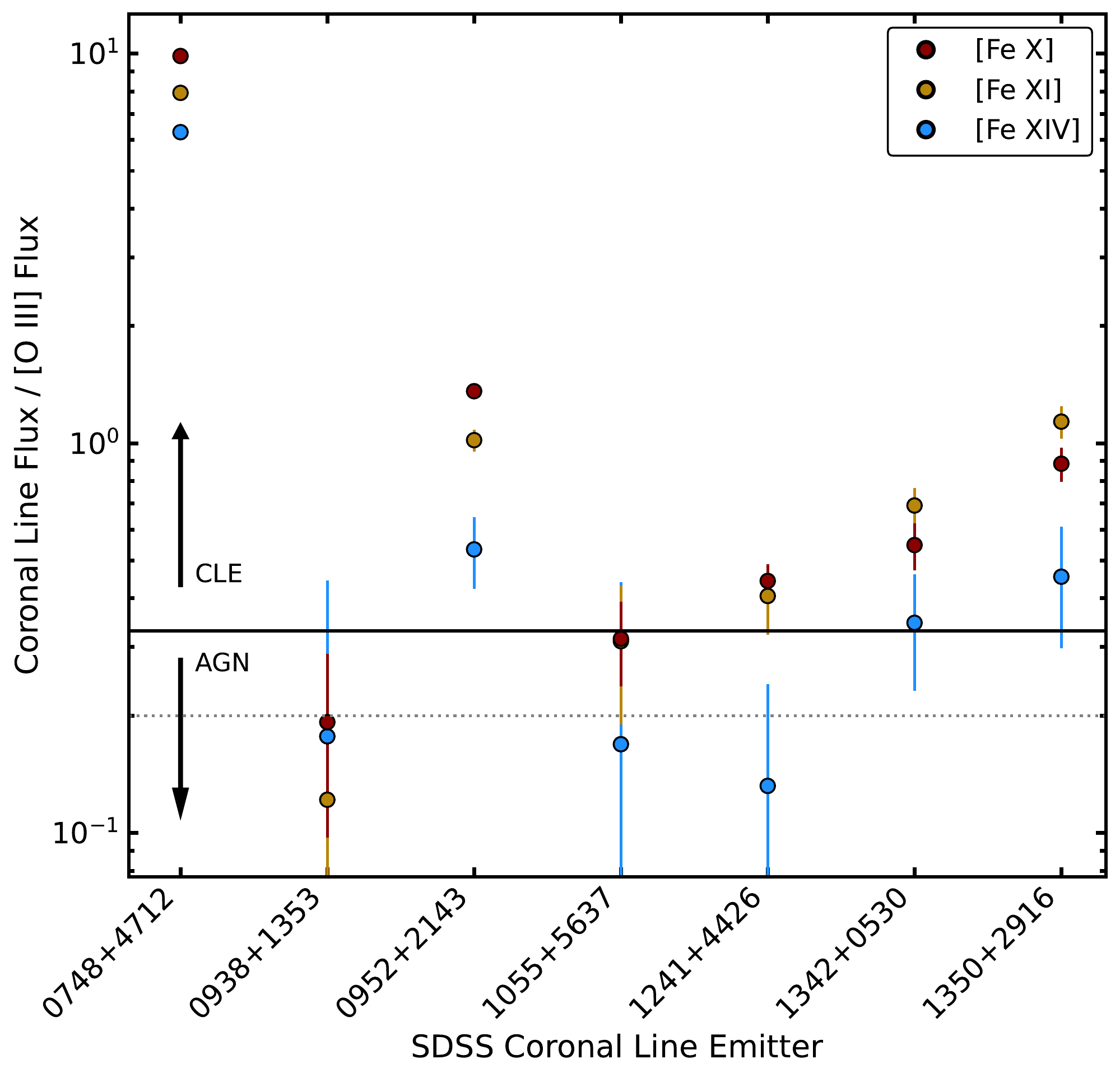}\hfill
 \caption{Ratio of the flux of [\ion{Fe}{x}], [\ion{Fe}{xi}] and [\ion{Fe}{xiv}] coronal lines to [\ion{O}{iii}] flux shown for each spectroscopically-selected CLE. The horizontal dotted line is the 20\% cutoff originally imposed by \citet{wang12}, above which sources are CLEs and below which sources are most likely AGNs. The horizontal solid line is our proposed threshold of 1/3 that removes the two SDSS sources convincingly shown to be AGNs.}
 \label{fig:cl_ratios}
\end{figure}

To understand this selection better, we used the published emission line measurements of \citet{wang12} for the SDSS spectroscopically-selected sample to investigate the ratio of CL flux to [\ion{O}{iii}] flux. These results are shown for this sub-sample in Figure \ref{fig:cl_ratios}. For simplicity, we only looked at the [\ion{Fe}{x}], [\ion{Fe}{xi}] and [\ion{Fe}{xiv}], although a similar analysis can be done for any CL as was done in \citet{wang12}. We find that while the initial 20\% threshold was reasonable, this does allow the non-transient sources to be selected as CLEs. A threshold of a CL with a flux greater than or equal to 1/3 of the [\ion{O}{iii}] flux successfully excludes 0938+1353 and 1055+5637, at least for this set of CLs. We, therefore, propose a stricter flux-ratio cutoff of 1/3 rather than 1/5 in selecting CLEs for future spectroscopic studies. We note that ratios of other CLs may require different ratios to exclude AGNs robustly, as 0938+1353 would not have been selected based on these 3 Fe lines alone.

Interestingly, 1241+4426 is also included in the sample of sources without fading CLs in \citet{yang13} and our CIGALE fits indicate an AGN component comparable in luminosity to the stellar output, although the WISE colors indicate a transient event. In Figure \ref{fig:cl_ratios}, 1241+4426 has [\ion{Fe}{x}] and [\ion{Fe}{xi}] emission above this 1/3 threshold, but not [\ion{Fe}{xiv}] and some others have fluxes with uncertainties allowing ratios below 1/5. Given the fluxes of two CLs above a ratio of 1/3 and the declining WISE light curve, it is most likely that 1241+4426 is powered by a TDE rather than an AGN.

When applying such a CL flux ratio threshold to the photometrically-selected sample, we can see from Fig. \ref{fig:spectra} that ASASSN-18jd, AT2021dms, and AT2022upj each surpass this ratio as several CLs are visually stronger than [\ion{O}{iii}]. ATLAS17jrp is close to this ratio, with spectra shown in \citet{onori22} supporting the CLE classification. With a larger sample of CLEs, it will be possible to better probe the temporal evolution of CL properties and their ratios with respect to other strong narrow emission lines.

\section{Discussion and Conclusions} \label{conc}

We have presented comparisons between CLEs and TDEs host-galaxy properties and temporal evolution. In general, we find that the CLEs are consistent with TDEs occurring in gas-rich environments. This is supported by several key pieces of evidence.

All but two of the spectroscopically-selected CLEs show declining WISE $W1 - W2$ color curves, providing strong evidence for the transient nature of these events. Even 1241+4426, for which spectroscopic follow-up suggested the possibility of an AGN \citep{yang13}, shows a transient WISE color evolution. Additionally, all of the photometrically-selected and all but two of the spectroscopically-selected CLEs show WISE colors consistent with normal galaxies, after accounting for transient-heated nuclear dust. The only two sources which consistently show AGN-like WISE colors are 0938+1353 and 1055+5637, which are likely AGNs based on spectroscopic follow-up \citep{yang13}. We also examined the ratio of CL flux to [\ion{O}{iii}] flux for the spectroscopically-selected sources. The AGN-like sources 0938+1353 and 1055+5637 have CL-to-[\ion{O}{iii}] flux ratios below 1/3 in all of the Fe CLs we test in Fig. \ref{fig:cl_ratios}, whereas the transient generally CLEs do not. For that reason, we propose a stricter CL-to-[\ion{O}{iii}] flux ratio of 1/3 as opposed to the ratio of 1/5 used in \citet{wang12} for selecting CLEs in the future.

In terms of host properties, the SFRs of the CLEs lie on the high end of the TDE distribution, likely as a result of the gas-rich environments needed for CL production. Many CLEs reside in the green valley although several spectroscopically-selected CLEs are in star-forming galaxies. The stellar masses of the spectroscopically-selected CLEs are consistent with TDEs. For the photometrically-selected CLEs, the stellar masses lie at the very highest end of the TDE distribution, with ASASSN-18jd having a substantially higher mass. When comparing the CLE hosts to the host galaxies of large MIR flares, the so-called MIRONGs, we find that the MIRONG hosts have significantly larger masses than the spectroscopically-selected CLEs but only slightly larger masses than the photometrically-selected CLEs. The SFRs of MIRONG hosts overlaps with both classes of CLEs, but the population extends to higher SFRs. This may suggest a continuum of gas and dust content from the nearly dust-free nuclei of normal TDEs, to moderate nuclear gas and dust for the CLEs, and finally significant gas and dust for the MIRONGs. Such a physical scenario would naturally explain the presence of MIR flares in both the CLE and MIRONG populations but the general lack of optical counterparts to the MIRONGs \citep{jiang21a, wang22a}.

Our sample of CLEs and TDEs show similar blackbody luminosity, radius, and temperature evolution. Only in the case of ASASSN-18jd is the luminosity somewhat higher than seen for TDEs. As compared to other classes of transients, the CLEs lie remarkably close to TDEs in the peak-luminosity/decline-rate parameter space. Nevertheless, 2 of the 4 CLEs in our sample lie slightly below the relationship, but within $\sim 2-3\sigma$ given the intrinsic scatter in the relationship.

The similarities of CLEs and TDEs both in terms of the galaxies in which they occur and the evolution of the transient events themselves suggests that CLEs are driven by TDEs. This supports earlier assertions of this nature \citep[e.g.,][]{wang12, yang13, palaversa16} made with less information on the transient properties. With improved follow-up enabled by early detections with transient surveys, we can now firmly connect CLEs with TDEs in gas-rich environments.

Ongoing and future spectroscopic surveys such as SDSS-V \citep{kollmeier17}, the Dark Energy Spectroscopic Instrument \citep[DESI; ][]{desi16}, and the Large Sky Area Multi-Object Fiber Spectroscopic Telescope \citep[LAMOST; ][]{lamost12} will discover many more CLEs. Assuming the initial rate estimates of $\sim 1 \times 10^{-5}$ gal$^{-1}$ yr$^{-1}$ \citep{wang12}, such surveys will find between a few CLEs (for LAMOST) and up to hundreds of CLEs (for DESI) per year. As compared to the original SDSS spectroscopic survey from which the first sample of CLES was selected, significant advancements can be made if new CLE spectra can be identified in real-time so that multi-wavelength follow-up efforts \citep[e.g.,][]{yang13, palaversa16, dou16, dai20, wang22a} can be initiated to better investigate this currently-understudied class of nuclear transients.

\section*{Data availability}
	
The data underlying this article are publicly available on the NASA High Energy Astrophysics Science Archive Research Center (HEASARC), Sloan Digital Sky Survey (SDSS) Science Archive Server, and the NASA/IPAC Infrared Science Archive (IRSA).

\section*{Acknowledgements}

We thank the referee for helpful comments that have improved this work. We thank Nidia Morrell for conducting the Magellan-LDSS3 observations of AT2021dms, and Jack Neustadt for sharing reduced spectra of ASASSN-18jd. We also thank Alexa Anderson for helpful comments on the manuscript.

JTH is supported by NASA grant 80NSSC22K0127. BJS is supported by NSF grants AST-1920392 and AST-1911074. Support for T.W.-S.H. was provided by NASA through the NASA Hubble Fellowship grant HST-HF2-51458.001-A awarded by the Space Telescope Science Institute (STScI), which is operated by the Association of Universities for Research in Astronomy, Inc., for NASA, under contract NAS5-26555.

Funding for the SDSS and SDSS-II has been provided by the Alfred P. Sloan Foundation, the Participating Institutions, the National Science Foundation, the U.S. Department of Energy, the National Aeronautics and Space Administration, the Japanese Monbukagakusho, the Max Planck Society, and the Higher Education Funding Council for England. The SDSS Web Site is http://www.sdss.org/.

The SDSS is managed by the Astrophysical Research Consortium for the Participating Institutions. The Participating Institutions are the American Museum of Natural History, Astrophysical Institute Potsdam, University of Basel, University of Cambridge, Case Western Reserve University, University of Chicago, Drexel University, Fermilab, the Institute for Advanced Study, the Japan Participation Group, Johns Hopkins University, the Joint Institute for Nuclear Astrophysics, the Kavli Institute for Particle Astrophysics and Cosmology, the Korean Scientist Group, the Chinese Academy of Sciences (LAMOST), Los Alamos National Laboratory, the Max-Planck-Institute for Astronomy (MPIA), the Max-Planck-Institute for Astrophysics (MPA), New Mexico State University, Ohio State University, University of Pittsburgh, University of Portsmouth, Princeton University, the United States Naval Observatory, and the University of Washington.

SDSS-III is managed by the Astrophysical Research Consortium for the Participating Institutions of the SDSS-III Collaboration including the University of Arizona, the Brazilian Participation Group, Brookhaven National Laboratory, Carnegie Mellon University, University of Florida, the French Participation Group, the German Participation Group, Harvard University, the Instituto de Astrofisica de Canarias, the Michigan State/Notre Dame/JINA Participation Group, Johns Hopkins University, Lawrence Berkeley National Laboratory, Max Planck Institute for Astrophysics, Max Planck Institute for Extraterrestrial Physics, New Mexico State University, New York University, Ohio State University, Pennsylvania State University, University of Portsmouth, Princeton University, the Spanish Participation Group, University of Tokyo, University of Utah, Vanderbilt University, University of Virginia, University of Washington, and Yale University.

This research has made use of the NASA/IPAC Infrared Science Archive, which is funded by the National Aeronautics and Space Administration and operated by the California Institute of Technology. We acknowledge the usage of the HyperLeda database (http://leda.univ-lyon1.fr). This paper includes data gathered with the 6.5 meter Magellan Telescopes located at Las Campanas Observatory, Chile.

\bibliography{bibliography}

\begin{thebibliography}{}
\makeatletter
\relax
\def\mn@urlcharsother{\let\do\@makeother \do\$\do\&\do\#\do\^\do\_\do\%\do\~}
\def\mn@doi{\begingroup\mn@urlcharsother \@ifnextchar [ {\mn@doi@}
  {\mn@doi@[]}}
\def\mn@doi@[#1]#2{\def\@tempa{#1}\ifx\@tempa\@empty \href
  {http://dx.doi.org/#2} {doi:#2}\else \href {http://dx.doi.org/#2} {#1}\fi
  \endgroup}
\def\mn@eprint#1#2{\mn@eprint@#1:#2::\@nil}
\def\mn@eprint@arXiv#1{\href {http://arxiv.org/abs/#1} {{\tt arXiv:#1}}}
\def\mn@eprint@dblp#1{\href {http://dblp.uni-trier.de/rec/bibtex/#1.xml}
  {dblp:#1}}
\def\mn@eprint@#1:#2:#3:#4\@nil{\def\@tempa {#1}\def\@tempb {#2}\def\@tempc
  {#3}\ifx \@tempc \@empty \let \@tempc \@tempb \let \@tempb \@tempa \fi \ifx
  \@tempb \@empty \def\@tempb {arXiv}\fi \@ifundefined
  {mn@eprint@\@tempb}{\@tempb:\@tempc}{\expandafter \expandafter \csname
  mn@eprint@\@tempb\endcsname \expandafter{\@tempc}}}

\bibitem[\protect\citeauthoryear{{Aguado} et~al.,}{{Aguado}
  et~al.}{2019}]{aguado19}
{Aguado} D.~S.,  et~al., 2019, \mn@doi [\apjs] {10.3847/1538-4365/aaf651},
  \href {https://ui.adsabs.harvard.edu/abs/2019ApJS..240...23A} {240, 23}

\bibitem[\protect\citeauthoryear{{Assef} et~al.,}{{Assef}
  et~al.}{2010}]{assef10}
{Assef} R.~J.,  et~al., 2010, \mn@doi [\apj] {10.1088/0004-637X/713/2/970},
  \href {https://ui.adsabs.harvard.edu/abs/2010ApJ...713..970A} {713, 970}

\bibitem[\protect\citeauthoryear{{Assef} et~al.,}{{Assef}
  et~al.}{2013}]{assef13}
{Assef} R.~J.,  et~al., 2013, \mn@doi [\apj] {10.1088/0004-637X/772/1/26},
  \href {http://adsabs.harvard.edu/abs/2013ApJ...772...26A} {772, 26}

\bibitem[\protect\citeauthoryear{{Auchettl}, {Guillochon}  \&
  {Ramirez-Ruiz}}{{Auchettl} et~al.}{2017}]{auchettl17}
{Auchettl} K.,  {Guillochon} J.,   {Ramirez-Ruiz} E.,  2017, \mn@doi [\apj]
  {10.3847/1538-4357/aa633b}, \href
  {http://adsabs.harvard.edu/abs/2017ApJ...838..149A} {838, 149}

\bibitem[\protect\citeauthoryear{{Auchettl}, {Ramirez-Ruiz}  \&
  {Guillochon}}{{Auchettl} et~al.}{2018}]{auchettl18}
{Auchettl} K.,  {Ramirez-Ruiz} E.,   {Guillochon} J.,  2018, \mn@doi [\apj]
  {10.3847/1538-4357/aa9b7c}, \href
  {http://adsabs.harvard.edu/abs/2018ApJ...852...37A} {852, 37}

\bibitem[\protect\citeauthoryear{{Barbarino}, {Carracedo}, {Tartaglia}  \&
  {Yaron}}{{Barbarino} et~al.}{2019}]{barbarino19}
{Barbarino} C.,  {Carracedo} A.~S.,  {Tartaglia} L.,   {Yaron} O.,  2019,
  Transient Name Server Classification Report, \href
  {https://ui.adsabs.harvard.edu/abs/2019TNSCR.287....1B} {2019-287, 1}

\bibitem[\protect\citeauthoryear{{Bennett}, {Larson}, {Weiland}  \&
  {Hinshaw}}{{Bennett} et~al.}{2014}]{bennett14}
{Bennett} C.~L.,  {Larson} D.,  {Weiland} J.~L.,   {Hinshaw} G.,  2014, \mn@doi
  [\apj] {10.1088/0004-637X/794/2/135}, \href
  {https://ui.adsabs.harvard.edu/abs/2014ApJ...794..135B} {794, 135}

\bibitem[\protect\citeauthoryear{{Blagorodnova} et~al.,}{{Blagorodnova}
  et~al.}{2017}]{blagorodnova17}
{Blagorodnova} N.,  et~al., 2017, \mn@doi [\apj] {10.3847/1538-4357/aa7579},
  \href {http://adsabs.harvard.edu/abs/2017ApJ...844...46B} {844, 46}

\bibitem[\protect\citeauthoryear{{Blagorodnova} et~al.,}{{Blagorodnova}
  et~al.}{2018}]{blagorodnova18}
{Blagorodnova} N.,  et~al., 2018, arXiv e-prints, \href
  {http://adsabs.harvard.edu/abs/2018arXiv180907446B} {}

\bibitem[\protect\citeauthoryear{{Boquien}, {Burgarella}, {Roehlly}, {Buat},
  {Ciesla}, {Corre}, {Inoue}  \& {Salas}}{{Boquien} et~al.}{2019}]{boquien19}
{Boquien} M.,  {Burgarella} D.,  {Roehlly} Y.,  {Buat} V.,  {Ciesla} L.,
  {Corre} D.,  {Inoue} A.~K.,   {Salas} H.,  2019, \mn@doi [\aap]
  {10.1051/0004-6361/201834156}, \href
  {https://ui.adsabs.harvard.edu/abs/2019A&A...622A.103B} {622, A103}

\bibitem[\protect\citeauthoryear{{Brinchmann}, {Charlot}, {White}, {Tremonti},
  {Kauffmann}, {Heckman}  \& {Brinkmann}}{{Brinchmann}
  et~al.}{2004}]{brinchmann04}
{Brinchmann} J.,  {Charlot} S.,  {White} S.~D.~M.,  {Tremonti} C.,  {Kauffmann}
  G.,  {Heckman} T.,   {Brinkmann} J.,  2004, \mn@doi [\mnras]
  {10.1111/j.1365-2966.2004.07881.x}, \href
  {https://ui.adsabs.harvard.edu/abs/2004MNRAS.351.1151B} {351, 1151}

\bibitem[\protect\citeauthoryear{{Bruzual} \& {Charlot}}{{Bruzual} \&
  {Charlot}}{2003}]{bruzual03}
{Bruzual} G.,  {Charlot} S.,  2003, \mn@doi [\mnras]
  {10.1046/j.1365-8711.2003.06897.x}, \href
  {http://adsabs.harvard.edu/abs/2003MNRAS.344.1000B} {344, 1000}

\bibitem[\protect\citeauthoryear{{Burgarella}, {Buat}  \&
  {Iglesias-P{\'a}ramo}}{{Burgarella} et~al.}{2005}]{burgarella05}
{Burgarella} D.,  {Buat} V.,   {Iglesias-P{\'a}ramo} J.,  2005, \mn@doi
  [\mnras] {10.1111/j.1365-2966.2005.09131.x}, \href
  {https://ui.adsabs.harvard.edu/abs/2005MNRAS.360.1413B} {360, 1413}

\bibitem[\protect\citeauthoryear{{Cardelli}, {Clayton}  \& {Mathis}}{{Cardelli}
  et~al.}{1989}]{cardelli89}
{Cardelli} J.~A.,  {Clayton} G.~C.,   {Mathis} J.~S.,  1989, \mn@doi [\apj]
  {10.1086/167900}, \href {http://adsabs.harvard.edu/abs/1989ApJ...345..245C}
  {345, 245}

\bibitem[\protect\citeauthoryear{{Cerqueira-Campos}, {Rodr{\'\i}guez-Ardila},
  {Riffel}, {Marinello}, {Prieto}  \& {Dahmer-Hahn}}{{Cerqueira-Campos}
  et~al.}{2021}]{cerqueiracampos21}
{Cerqueira-Campos} F.~C.,  {Rodr{\'\i}guez-Ardila} A.,  {Riffel} R.,
  {Marinello} M.,  {Prieto} A.,   {Dahmer-Hahn} L.~G.,  2021, \mn@doi [\mnras]
  {10.1093/mnras/staa3320}, \href
  {https://ui.adsabs.harvard.edu/abs/2021MNRAS.500.2666C} {500, 2666}

\bibitem[\protect\citeauthoryear{{Chambers} et~al.,}{{Chambers}
  et~al.}{2016}]{chambers16}
{Chambers} K.~C.,  et~al., 2016, preprint, \href
  {http://adsabs.harvard.edu/abs/2016arXiv161205560C} {} (\mn@eprint {arXiv}
  {1612.05560})

\bibitem[\protect\citeauthoryear{{Cui} et~al.,}{{Cui} et~al.}{2012}]{lamost12}
{Cui} X.-Q.,  et~al., 2012, \mn@doi [Research in Astronomy and Astrophysics]
  {10.1088/1674-4527/12/9/003}, \href
  {https://ui.adsabs.harvard.edu/abs/2012RAA....12.1197C} {12, 1197}

\bibitem[\protect\citeauthoryear{{DESI Collaboration} et~al.,}{{DESI
  Collaboration} et~al.}{2016}]{desi16}
{DESI Collaboration} et~al., 2016, \mn@doi [arXiv e-prints]
  {10.48550/arXiv.1611.00036}, \href
  {https://ui.adsabs.harvard.edu/abs/2016arXiv161100036D} {p. arXiv:1611.00036}

\bibitem[\protect\citeauthoryear{{Dai}, {Shu}, {Jiang}, {Dou}, {Liu}, {Yang},
  {Zhang}  \& {Wang}}{{Dai} et~al.}{2020}]{dai20}
{Dai} B.~B.,  {Shu} X.~W.,  {Jiang} N.,  {Dou} L.~M.,  {Liu} D.~Z.,  {Yang}
  C.~W.,  {Zhang} F.~B.,   {Wang} T.~G.,  2020, \mn@doi [\apjl]
  {10.3847/2041-8213/ab97ac}, \href
  {https://ui.adsabs.harvard.edu/abs/2020ApJ...896L..27D} {896, L27}

\bibitem[\protect\citeauthoryear{{Dou}, {Wang}, {Jiang}, {Yang}, {Lyu}  \&
  {Zhou}}{{Dou} et~al.}{2016}]{dou16}
{Dou} L.,  {Wang} T.-g.,  {Jiang} N.,  {Yang} C.,  {Lyu} J.,   {Zhou} H.,
  2016, \mn@doi [\apj] {10.3847/0004-637X/832/2/188}, \href
  {https://ui.adsabs.harvard.edu/abs/2016ApJ...832..188D} {832, 188}

\bibitem[\protect\citeauthoryear{{Draine} \& {Lee}}{{Draine} \&
  {Lee}}{1984}]{draine84}
{Draine} B.~T.,  {Lee} H.~M.,  1984, \mn@doi [\apj] {10.1086/162480}, \href
  {https://ui.adsabs.harvard.edu/abs/1984ApJ...285...89D} {285, 89}

\bibitem[\protect\citeauthoryear{{Eisenstein} et~al.,}{{Eisenstein}
  et~al.}{2011}]{eisenstein11}
{Eisenstein} D.~J.,  et~al., 2011, \mn@doi [\aj] {10.1088/0004-6256/142/3/72},
  \href {https://ui.adsabs.harvard.edu/abs/2011AJ....142...72E} {142, 72}

\bibitem[\protect\citeauthoryear{{Evans} \& {Kochanek}}{{Evans} \&
  {Kochanek}}{1989}]{evans89}
{Evans} C.~R.,  {Kochanek} C.~S.,  1989, \mn@doi [\apjl] {10.1086/185567},
  \href {http://adsabs.harvard.edu/abs/1989ApJ...346L..13E} {346, L13}

\bibitem[\protect\citeauthoryear{{Flewelling} et~al.,}{{Flewelling}
  et~al.}{2016}]{flewelling16}
{Flewelling} H.~A.,  et~al., 2016, preprint, \href
  {http://adsabs.harvard.edu/abs/2016arXiv161205243F} {} (\mn@eprint {arXiv}
  {1612.05243})

\bibitem[\protect\citeauthoryear{{Forster} et~al.,}{{Forster}
  et~al.}{2021}]{forster21}
{Forster} F.,  et~al., 2021, Transient Name Server Discovery Report, \href
  {https://ui.adsabs.harvard.edu/abs/2021TNSTR.542....1F} {2021-542, 1}

\bibitem[\protect\citeauthoryear{{French}}{{French}}{2021}]{french21}
{French} K.~D.,  2021, \mn@doi [\pasp] {10.1088/1538-3873/ac0a59}, \href
  {https://ui.adsabs.harvard.edu/abs/2021PASP..133g2001F} {133, 072001}

\bibitem[\protect\citeauthoryear{{French}, {Arcavi}  \& {Zabludoff}}{{French}
  et~al.}{2016}]{french16}
{French} K.~D.,  {Arcavi} I.,   {Zabludoff} A.,  2016, \mn@doi [\apjl]
  {10.3847/2041-8205/818/1/L21}, \href
  {http://adsabs.harvard.edu/abs/2016ApJ...818L..21F} {818, L21}

\bibitem[\protect\citeauthoryear{{French}, {Arcavi}  \& {Zabludoff}}{{French}
  et~al.}{2017}]{french17}
{French} K.~D.,  {Arcavi} I.,   {Zabludoff} A.,  2017, \mn@doi [\apj]
  {10.3847/1538-4357/835/2/176}, \href
  {https://ui.adsabs.harvard.edu/abs/2017ApJ...835..176F} {835, 176}

\bibitem[\protect\citeauthoryear{{French}, {Arcavi}, {Zabludoff}, {Stone},
  {Hiramatsu}, {van Velzen}, {McCully}  \& {Jiang}}{{French}
  et~al.}{2020}]{french20}
{French} K.~D.,  {Arcavi} I.,  {Zabludoff} A.~I.,  {Stone} N.,  {Hiramatsu} D.,
   {van Velzen} S.,  {McCully} C.,   {Jiang} N.,  2020, \mn@doi [\apj]
  {10.3847/1538-4357/ab7450}, \href
  {https://ui.adsabs.harvard.edu/abs/2020ApJ...891...93F} {891, 93}

\bibitem[\protect\citeauthoryear{{Fulton} et~al.,}{{Fulton}
  et~al.}{2022}]{fulton22}
{Fulton} M.,  et~al., 2022, Transient Name Server Classification Report, \href
  {https://ui.adsabs.harvard.edu/abs/2022TNSCR3245....1F} {2022-3245, 1}

\bibitem[\protect\citeauthoryear{{Gehrels} et~al.,}{{Gehrels}
  et~al.}{2004}]{gehrels04}
{Gehrels} N.,  et~al., 2004, \mn@doi [\apj] {10.1086/422091}, \href
  {http://adsabs.harvard.edu/abs/2004ApJ...611.1005G} {611, 1005}

\bibitem[\protect\citeauthoryear{{Gezari}}{{Gezari}}{2021}]{gezari21}
{Gezari} S.,  2021, \mn@doi [\araa] {10.1146/annurev-astro-111720-030029},
  \href {https://ui.adsabs.harvard.edu/abs/2021ARA&A..59...21G} {59, 21}

\bibitem[\protect\citeauthoryear{{Gezari} et~al.,}{{Gezari}
  et~al.}{2012}]{gezari12b}
{Gezari} S.,  et~al., 2012, \mn@doi [\nat] {10.1038/nature10990}, \href
  {http://adsabs.harvard.edu/abs/2012Natur.485..217G} {485, 217}

\bibitem[\protect\citeauthoryear{{Hammerstein} et~al.,}{{Hammerstein}
  et~al.}{2021}]{hammerstein21}
{Hammerstein} E.,  et~al., 2021, \mn@doi [\apjl] {10.3847/2041-8213/abdcb4},
  \href {https://ui.adsabs.harvard.edu/abs/2021ApJ...908L..20H} {908, L20}

\bibitem[\protect\citeauthoryear{{Hammerstein} et~al.,}{{Hammerstein}
  et~al.}{2023}]{hammerstein23}
{Hammerstein} E.,  et~al., 2023, \mn@doi [\apj] {10.3847/1538-4357/aca283},
  \href {https://ui.adsabs.harvard.edu/abs/2023ApJ...942....9H} {942, 9}

\bibitem[\protect\citeauthoryear{{Hildebrand}}{{Hildebrand}}{1983}]{hildebrand83}
{Hildebrand} R.~H.,  1983, \qjras, \href
  {https://ui.adsabs.harvard.edu/abs/1983QJRAS..24..267H} {24, 267}

\bibitem[\protect\citeauthoryear{{Hinkle}}{{Hinkle}}{2022}]{hinkle22b}
{Hinkle} J.~T.,  2022, \mn@doi [arXiv e-prints] {10.48550/arXiv.2210.15681},
  \href {https://ui.adsabs.harvard.edu/abs/2022arXiv221015681H} {p.
  arXiv:2210.15681}

\bibitem[\protect\citeauthoryear{{Hinkle}, {Holoien}, {Shappee}, {Auchettl},
  {Kochanek}, {Stanek}, {Payne}  \& {Thompson}}{{Hinkle}
  et~al.}{2020}]{hinkle20a}
{Hinkle} J.~T.,  {Holoien} T. W.~S.,  {Shappee} B.~J.,  {Auchettl} K.,
  {Kochanek} C.~S.,  {Stanek} K.~Z.,  {Payne} A.~V.,   {Thompson} T.~A.,  2020,
  arXiv e-prints, \href {https://ui.adsabs.harvard.edu/abs/2020arXiv200108215H}
  {p. arXiv:2001.08215}

\bibitem[\protect\citeauthoryear{{Hinkle} et~al.,}{{Hinkle}
  et~al.}{2021a}]{hinkle21a}
{Hinkle} J.~T.,  et~al., 2021a, \mn@doi [\mnras] {10.1093/mnras/staa3170},
  \href {https://ui.adsabs.harvard.edu/abs/2021MNRAS.500.1673H} {500, 1673}

\bibitem[\protect\citeauthoryear{{Hinkle}, {Holoien}, {Shappee}  \&
  {Auchettl}}{{Hinkle} et~al.}{2021b}]{hinkle21b}
{Hinkle} J.~T.,  {Holoien} T. W.~S.,  {Shappee} B.~J.,   {Auchettl} K.,  2021b,
  \mn@doi [\apj] {10.3847/1538-4357/abe4d8}, \href
  {https://ui.adsabs.harvard.edu/abs/2021ApJ...910...83H} {910, 83}

\bibitem[\protect\citeauthoryear{{Hinkle} et~al.,}{{Hinkle}
  et~al.}{2022}]{hinkle22a}
{Hinkle} J.~T.,  et~al., 2022, \mn@doi [\apj] {10.3847/1538-4357/ac5f54}, \href
  {https://ui.adsabs.harvard.edu/abs/2022ApJ...930...12H} {930, 12}

\bibitem[\protect\citeauthoryear{{Ho}}{{Ho}}{2008}]{ho08}
{Ho} L.~C.,  2008, \mn@doi [\araa] {10.1146/annurev.astro.45.051806.110546},
  \href {https://ui.adsabs.harvard.edu/abs/2008ARA&A..46..475H} {46, 475}

\bibitem[\protect\citeauthoryear{{Holoien} et~al.,}{{Holoien}
  et~al.}{2014}]{holoien14b}
{Holoien} T.~W.-S.,  et~al., 2014, \mn@doi [\mnras] {10.1093/mnras/stu1922},
  \href {http://adsabs.harvard.edu/abs/2014MNRAS.445.3263H} {445, 3263}

\bibitem[\protect\citeauthoryear{{Holoien} et~al.,}{{Holoien}
  et~al.}{2016a}]{holoien16a}
{Holoien} T.~W.-S.,  et~al., 2016a, \mn@doi [\mnras] {10.1093/mnras/stv2486},
  \href {http://adsabs.harvard.edu/abs/2016MNRAS.455.2918H} {455, 2918}

\bibitem[\protect\citeauthoryear{{Holoien} et~al.,}{{Holoien}
  et~al.}{2016b}]{holoien16b}
{Holoien} T.~W.-S.,  et~al., 2016b, \mn@doi [\mnras] {10.1093/mnras/stw2272},
  \href {http://adsabs.harvard.edu/abs/2016MNRAS.463.3813H} {463, 3813}

\bibitem[\protect\citeauthoryear{{Holoien} et~al.,}{{Holoien}
  et~al.}{2019a}]{holoien19b}
{Holoien} T.~W.~S.,  et~al., 2019a, \mn@doi [\apj] {10.3847/1538-4357/ab2ae1},
  \href {https://ui.adsabs.harvard.edu/abs/2019ApJ...880..120H} {880, 120}

\bibitem[\protect\citeauthoryear{{Holoien} et~al.,}{{Holoien}
  et~al.}{2019b}]{holoien19c}
{Holoien} T. W.~S.,  et~al., 2019b, \mn@doi [\apj] {10.3847/1538-4357/ab3c66},
  \href {https://ui.adsabs.harvard.edu/abs/2019ApJ...883..111H} {883, 111}

\bibitem[\protect\citeauthoryear{{Holoien} et~al.,}{{Holoien}
  et~al.}{2020}]{holoien20}
{Holoien} T. W.~S.,  et~al., 2020, arXiv e-prints, \href
  {https://ui.adsabs.harvard.edu/abs/2020arXiv200313693H} {p. arXiv:2003.13693}

\bibitem[\protect\citeauthoryear{{Hung} et~al.,}{{Hung} et~al.}{2017}]{hung17}
{Hung} T.,  et~al., 2017, \mn@doi [\apj] {10.3847/1538-4357/aa7337}, \href
  {http://adsabs.harvard.edu/abs/2017ApJ...842...29H} {842, 29}

\bibitem[\protect\citeauthoryear{{Jiang} et~al.,}{{Jiang}
  et~al.}{2021a}]{jiang21a}
{Jiang} N.,  et~al., 2021a, \mn@doi [\apjs] {10.3847/1538-4365/abd1dc}, \href
  {https://ui.adsabs.harvard.edu/abs/2021ApJS..252...32J} {252, 32}

\bibitem[\protect\citeauthoryear{{Jiang}, {Wang}, {Hu}, {Sun}, {Dou}  \&
  {Xiao}}{{Jiang} et~al.}{2021b}]{jiang21b}
{Jiang} N.,  {Wang} T.,  {Hu} X.,  {Sun} L.,  {Dou} L.,   {Xiao} L.,  2021b,
  \mn@doi [\apj] {10.3847/1538-4357/abe772}, \href
  {https://ui.adsabs.harvard.edu/abs/2021ApJ...911...31J} {911, 31}

\bibitem[\protect\citeauthoryear{{Kollmeier} et~al.,}{{Kollmeier}
  et~al.}{2017}]{kollmeier17}
{Kollmeier} J.~A.,  et~al., 2017, \mn@doi [arXiv e-prints]
  {10.48550/arXiv.1711.03234}, \href
  {https://ui.adsabs.harvard.edu/abs/2017arXiv171103234K} {p. arXiv:1711.03234}

\bibitem[\protect\citeauthoryear{{Komossa} et~al.,}{{Komossa}
  et~al.}{2008}]{komossa08}
{Komossa} S.,  et~al., 2008, \mn@doi [\apjl] {10.1086/588281}, \href
  {https://ui.adsabs.harvard.edu/abs/2008ApJ...678L..13K} {678, L13}

\bibitem[\protect\citeauthoryear{{Komossa} et~al.,}{{Komossa}
  et~al.}{2009}]{komossa09}
{Komossa} S.,  et~al., 2009, \mn@doi [\apj] {10.1088/0004-637X/701/1/105},
  \href {https://ui.adsabs.harvard.edu/abs/2009ApJ...701..105K} {701, 105}

\bibitem[\protect\citeauthoryear{{Kool} et~al.,}{{Kool} et~al.}{2020}]{kool20}
{Kool} E.~C.,  et~al., 2020, \mn@doi [\mnras] {10.1093/mnras/staa2351}, \href
  {https://ui.adsabs.harvard.edu/abs/2020MNRAS.498.2167K} {498, 2167}

\bibitem[\protect\citeauthoryear{{Kormendy} \& {Ho}}{{Kormendy} \&
  {Ho}}{2013}]{kormendy13}
{Kormendy} J.,  {Ho} L.~C.,  2013, \mn@doi [\araa]
  {10.1146/annurev-astro-082708-101811}, \href
  {https://ui.adsabs.harvard.edu/abs/2013ARA&A..51..511K} {51, 511}

\bibitem[\protect\citeauthoryear{{Kormendy} \& {Richstone}}{{Kormendy} \&
  {Richstone}}{1995}]{kormendy95}
{Kormendy} J.,  {Richstone} D.,  1995, \mn@doi [\araa]
  {10.1146/annurev.aa.33.090195.003053}, \href
  {https://ui.adsabs.harvard.edu/abs/1995ARA&A..33..581K} {33, 581}

\bibitem[\protect\citeauthoryear{{Kriek}, {van Dokkum}, {Labb{\'e}}, {Franx},
  {Illingworth}, {Marchesini}  \& {Quadri}}{{Kriek} et~al.}{2009}]{kriek09}
{Kriek} M.,  {van Dokkum} P.~G.,  {Labb{\'e}} I.,  {Franx} M.,  {Illingworth}
  G.~D.,  {Marchesini} D.,   {Quadri} R.~F.,  2009, \mn@doi [\apj]
  {10.1088/0004-637X/700/1/221}, \href
  {http://adsabs.harvard.edu/abs/2009ApJ...700..221K} {700, 221}

\bibitem[\protect\citeauthoryear{{Leloudas} et~al.,}{{Leloudas}
  et~al.}{2019}]{leloudas19}
{Leloudas} G.,  et~al., 2019, arXiv e-prints, \href
  {http://adsabs.harvard.edu/abs/2019arXiv190303120L} {}

\bibitem[\protect\citeauthoryear{{Lu}, {Kumar}  \& {Evans}}{{Lu}
  et~al.}{2016}]{lu16}
{Lu} W.,  {Kumar} P.,   {Evans} N.~J.,  2016, \mn@doi [\mnras]
  {10.1093/mnras/stw307}, \href
  {https://ui.adsabs.harvard.edu/abs/2016MNRAS.458..575L} {458, 575}

\bibitem[\protect\citeauthoryear{{Magorrian} et~al.,}{{Magorrian}
  et~al.}{1998}]{magorrian98}
{Magorrian} J.,  et~al., 1998, \mn@doi [\aj] {10.1086/300353}, \href
  {https://ui.adsabs.harvard.edu/abs/1998AJ....115.2285M} {115, 2285}

\bibitem[\protect\citeauthoryear{{Mainzer} et~al.,}{{Mainzer}
  et~al.}{2014}]{mainzer14}
{Mainzer} A.,  et~al., 2014, \mn@doi [\apj] {10.1088/0004-637X/792/1/30}, \href
  {https://ui.adsabs.harvard.edu/abs/2014ApJ...792...30M} {792, 30}

\bibitem[\protect\citeauthoryear{{Makarov}, {Prugniel}, {Terekhova}, {Courtois}
   \& {Vauglin}}{{Makarov} et~al.}{2014}]{makarov14}
{Makarov} D.,  {Prugniel} P.,  {Terekhova} N.,  {Courtois} H.,   {Vauglin} I.,
  2014, \mn@doi [\aap] {10.1051/0004-6361/201423496}, \href
  {https://ui.adsabs.harvard.edu/abs/2014A&A...570A..13M} {570, A13}

\bibitem[\protect\citeauthoryear{{Martin} et~al.,}{{Martin}
  et~al.}{2005}]{martin05}
{Martin} D.~C.,  et~al., 2005, \mn@doi [\apjl] {10.1086/426387}, \href
  {https://ui.adsabs.harvard.edu/abs/2005ApJ...619L...1M} {619, L1}

\bibitem[\protect\citeauthoryear{{Mattila} et~al.,}{{Mattila}
  et~al.}{2018}]{mattila18}
{Mattila} S.,  et~al., 2018, \mn@doi [Science] {10.1126/science.aao4669}, \href
  {https://ui.adsabs.harvard.edu/abs/2018Sci...361..482M} {361, 482}

\bibitem[\protect\citeauthoryear{{Negus}, {Comerford}, {M{\"u}ller
  S{\'a}nchez}, {Barrera-Ballesteros}, {Drory}, {Rembold}  \& {Riffel}}{{Negus}
  et~al.}{2021}]{negus21}
{Negus} J.,  {Comerford} J.~M.,  {M{\"u}ller S{\'a}nchez} F.,
  {Barrera-Ballesteros} J.~K.,  {Drory} N.,  {Rembold} S.~B.,   {Riffel} R.~A.,
   2021, \mn@doi [\apj] {10.3847/1538-4357/ac1343}, \href
  {https://ui.adsabs.harvard.edu/abs/2021ApJ...920...62N} {920, 62}

\bibitem[\protect\citeauthoryear{{Neustadt} et~al.,}{{Neustadt}
  et~al.}{2020}]{neustadt20}
{Neustadt} J.~M.~M.,  et~al., 2020, \mn@doi [\mnras] {10.1093/mnras/staa859},
  \href {https://ui.adsabs.harvard.edu/abs/2020MNRAS.494.2538N} {494, 2538}

\bibitem[\protect\citeauthoryear{{Newsome}, {Dgany}, {Arcavi}, {Howell},
  {McCully}, {Gonzalez}  \& {Pellegrino}}{{Newsome} et~al.}{2022}]{newsome22}
{Newsome} M.,  {Dgany} Y.,  {Arcavi} I.,  {Howell} D.~A.,  {McCully} C.,
  {Gonzalez} E.~P.,   {Pellegrino} C.,  2022, Transient Name Server
  Classification Report, \href
  {https://ui.adsabs.harvard.edu/abs/2022TNSCR3231....1N} {2022-3231, 1}

\bibitem[\protect\citeauthoryear{{Nicholl} et~al.,}{{Nicholl}
  et~al.}{2020}]{nicholl20}
{Nicholl} M.,  et~al., 2020, arXiv e-prints, \href
  {https://ui.adsabs.harvard.edu/abs/2020arXiv200602454N} {p. arXiv:2006.02454}

\bibitem[\protect\citeauthoryear{{Onori} et~al.,}{{Onori}
  et~al.}{2022}]{onori22}
{Onori} F.,  et~al., 2022, \mn@doi [\mnras] {10.1093/mnras/stac2673}, \href
  {https://ui.adsabs.harvard.edu/abs/2022MNRAS.517...76O} {517, 76}

\bibitem[\protect\citeauthoryear{{Palaversa}, {Gezari}, {Sesar}, {Stuart},
  {Wozniak}, {Holl}  \& {Ivezi{\'c}}}{{Palaversa} et~al.}{2016}]{palaversa16}
{Palaversa} L.,  {Gezari} S.,  {Sesar} B.,  {Stuart} J.~S.,  {Wozniak} P.,
  {Holl} B.,   {Ivezi{\'c}} {\v{Z}}.,  2016, \mn@doi [\apj]
  {10.3847/0004-637X/819/2/151}, \href
  {https://ui.adsabs.harvard.edu/abs/2016ApJ...819..151P} {819, 151}

\bibitem[\protect\citeauthoryear{{Phinney}}{{Phinney}}{1989}]{phinney89}
{Phinney} E.~S.,  1989, \mn@doi [\nat] {10.1038/340595a0}, \href
  {http://adsabs.harvard.edu/abs/1989Natur.340..595P} {340, 595}

\bibitem[\protect\citeauthoryear{{Poole} et~al.,}{{Poole}
  et~al.}{2008}]{poole08}
{Poole} T.~S.,  et~al., 2008, \mn@doi [\mnras]
  {10.1111/j.1365-2966.2007.12563.x}, \href
  {http://adsabs.harvard.edu/abs/2008MNRAS.383..627P} {383, 627}

\bibitem[\protect\citeauthoryear{{Prieto}, {Rodr{\'\i}guez-Ardila}, {Panda}  \&
  {Marinello}}{{Prieto} et~al.}{2022}]{prieto22}
{Prieto} A.,  {Rodr{\'\i}guez-Ardila} A.,  {Panda} S.,   {Marinello} M.,  2022,
  \mn@doi [\mnras] {10.1093/mnras/stab3414}, \href
  {https://ui.adsabs.harvard.edu/abs/2022MNRAS.510.1010P} {510, 1010}

\bibitem[\protect\citeauthoryear{{Rees}}{{Rees}}{1988}]{rees88}
{Rees} M.~J.,  1988, \mn@doi [\nat] {10.1038/333523a0}, \href
  {http://adsabs.harvard.edu/abs/1988Natur.333..523R} {333, 523}

\bibitem[\protect\citeauthoryear{{Rodrigo}, {Solano}  \& {Bayo}}{{Rodrigo}
  et~al.}{2012}]{rodrigo12}
{Rodrigo} C.,  {Solano} E.,   {Bayo} A.,  2012, {SVO Filter Profile Service
  Version 1.0}, IVOA Working Draft 15 October 2012,
  \mn@doi{10.5479/ADS/bib/2012ivoa.rept.1015R}

\bibitem[\protect\citeauthoryear{{Roming} et~al.,}{{Roming}
  et~al.}{2005}]{roming05}
{Roming} P.~W.~A.,  et~al., 2005, \mn@doi [SSR] {10.1007/s11214-005-5095-4},
  \href {http://adsabs.harvard.edu/abs/2005SSRv..120...95R} {120, 95}

\bibitem[\protect\citeauthoryear{{Salpeter}}{{Salpeter}}{1955}]{salpeter55}
{Salpeter} E.~E.,  1955, \mn@doi [\apj] {10.1086/145971}, \href
  {https://ui.adsabs.harvard.edu/abs/1955ApJ...121..161S} {121, 161}

\bibitem[\protect\citeauthoryear{{Sheng}, {Wang}, {Jiang}, {Yang}, {Yan}, {Dou}
   \& {Peng}}{{Sheng} et~al.}{2017}]{sheng17}
{Sheng} Z.,  {Wang} T.,  {Jiang} N.,  {Yang} C.,  {Yan} L.,  {Dou} L.,   {Peng}
  B.,  2017, \mn@doi [\apjl] {10.3847/2041-8213/aa85de}, \href
  {https://ui.adsabs.harvard.edu/abs/2017ApJ...846L...7S} {846, L7}

\bibitem[\protect\citeauthoryear{{Skrutskie} et~al.,}{{Skrutskie}
  et~al.}{2006}]{skrutskie06}
{Skrutskie} M.~F.,  et~al., 2006, \mn@doi [\aj] {10.1086/498708}, \href
  {http://adsabs.harvard.edu/abs/2006AJ....131.1163S} {131, 1163}

\bibitem[\protect\citeauthoryear{{Smartt} et~al.,}{{Smartt}
  et~al.}{2015}]{smartt15}
{Smartt} S.~J.,  et~al., 2015, \mn@doi [\aap] {10.1051/0004-6361/201425237},
  \href {http://adsabs.harvard.edu/abs/2015A%26A...579A..40S} {579, A40}

\bibitem[\protect\citeauthoryear{{Stalevski}, {Fritz}, {Baes}, {Nakos}  \&
  {Popovi{\'c}}}{{Stalevski} et~al.}{2012}]{stalevski12}
{Stalevski} M.,  {Fritz} J.,  {Baes} M.,  {Nakos} T.,   {Popovi{\'c}}
  L.~{\v{C}}.,  2012, \mn@doi [\mnras] {10.1111/j.1365-2966.2011.19775.x},
  \href {https://ui.adsabs.harvard.edu/abs/2012MNRAS.420.2756S} {420, 2756}

\bibitem[\protect\citeauthoryear{{Stalevski}, {Ricci}, {Ueda}, {Lira}, {Fritz}
  \& {Baes}}{{Stalevski} et~al.}{2016}]{stalevksi16}
{Stalevski} M.,  {Ricci} C.,  {Ueda} Y.,  {Lira} P.,  {Fritz} J.,   {Baes} M.,
  2016, \mn@doi [\mnras] {10.1093/mnras/stw444}, \href
  {https://ui.adsabs.harvard.edu/abs/2016MNRAS.458.2288S} {458, 2288}

\bibitem[\protect\citeauthoryear{{Stern} et~al.,}{{Stern}
  et~al.}{2004}]{stern04}
{Stern} D.,  et~al., 2004, \mn@doi [\apj] {10.1086/422744}, \href
  {http://adsabs.harvard.edu/abs/2004ApJ...612..690S} {612, 690}

\bibitem[\protect\citeauthoryear{{Stone} \& {van Velzen}}{{Stone} \& {van
  Velzen}}{2016}]{stone16b}
{Stone} N.~C.,  {van Velzen} S.,  2016, \mn@doi [\apjl]
  {10.3847/2041-8205/825/1/L14}, \href
  {https://ui.adsabs.harvard.edu/abs/2016ApJ...825L..14S} {825, L14}

\bibitem[\protect\citeauthoryear{{Tadhunter}, {Spence}, {Rose}, {Mullaney}  \&
  {Crowther}}{{Tadhunter} et~al.}{2017}]{tadhunter17}
{Tadhunter} C.,  {Spence} R.,  {Rose} M.,  {Mullaney} J.,   {Crowther} P.,
  2017, \mn@doi [Nature Astronomy] {10.1038/s41550-017-0061}, \href
  {https://ui.adsabs.harvard.edu/abs/2017NatAs...1E..61T} {1, 0061}

\bibitem[\protect\citeauthoryear{{Tody}}{{Tody}}{1986}]{tody86}
{Tody} D.,  1986, in {Crawford} D.~L.,  ed.,  Society of Photo-Optical
  Instrumentation Engineers (SPIE) Conference Series Vol. 627, Instrumentation
  in astronomy VI. p.~733, \mn@doi{10.1117/12.968154}

\bibitem[\protect\citeauthoryear{{Tody}}{{Tody}}{1993}]{tody93}
{Tody} D.,  1993, in {Hanisch} R.~J.,  {Brissenden} R.~J.~V.,   {Barnes} J.,
  eds,  Astronomical Society of the Pacific Conference Series Vol. 52,
  Astronomical Data Analysis Software and Systems II. p.~173

\bibitem[\protect\citeauthoryear{{Ulmer}}{{Ulmer}}{1999}]{ulmer99}
{Ulmer} A.,  1999, \mn@doi [\apj] {10.1086/306909}, \href
  {http://adsabs.harvard.edu/abs/1999ApJ...514..180U} {514, 180}

\bibitem[\protect\citeauthoryear{{Wang}, {Zhou}, {Wang}, {Lu}  \& {Xu}}{{Wang}
  et~al.}{2011}]{wang11}
{Wang} T.-G.,  {Zhou} H.-Y.,  {Wang} L.-F.,  {Lu} H.-L.,   {Xu} D.,  2011,
  \mn@doi [\apj] {10.1088/0004-637X/740/2/85}, \href
  {http://adsabs.harvard.edu/abs/2011ApJ...740...85W} {740, 85}

\bibitem[\protect\citeauthoryear{{Wang}, {Zhou}, {Komossa}, {Wang}, {Yuan}  \&
  {Yang}}{{Wang} et~al.}{2012}]{wang12}
{Wang} T.-G.,  {Zhou} H.-Y.,  {Komossa} S.,  {Wang} H.-Y.,  {Yuan} W.,   {Yang}
  C.,  2012, \mn@doi [\apj] {10.1088/0004-637X/749/2/115}, \href
  {https://ui.adsabs.harvard.edu/abs/2012ApJ...749..115W} {749, 115}

\bibitem[\protect\citeauthoryear{{Wang}, {Yan}, {Dou}, {Jiang}, {Sheng}  \&
  {Yang}}{{Wang} et~al.}{2018}]{wang18}
{Wang} T.,  {Yan} L.,  {Dou} L.,  {Jiang} N.,  {Sheng} Z.,   {Yang} C.,  2018,
  \mn@doi [\mnras] {10.1093/mnras/sty465}, \href
  {https://ui.adsabs.harvard.edu/abs/2018MNRAS.477.2943W} {477, 2943}

\bibitem[\protect\citeauthoryear{{Wang} et~al.,}{{Wang}
  et~al.}{2022a}]{wang22a}
{Wang} Y.,  et~al., 2022a, \mn@doi [\apjs] {10.3847/1538-4365/ac33a6}, \href
  {https://ui.adsabs.harvard.edu/abs/2022ApJS..258...21W} {258, 21}

\bibitem[\protect\citeauthoryear{{Wang} et~al.,}{{Wang}
  et~al.}{2022b}]{wang22b}
{Wang} Y.,  et~al., 2022b, \mn@doi [\apjl] {10.3847/2041-8213/ac6670}, \href
  {https://ui.adsabs.harvard.edu/abs/2022ApJ...930L...4W} {930, L4}

\bibitem[\protect\citeauthoryear{{Wevers} et~al.,}{{Wevers}
  et~al.}{2019}]{wevers19}
{Wevers} T.,  et~al., 2019, \mn@doi [\mnras] {10.1093/mnras/stz1976}, \href
  {https://ui.adsabs.harvard.edu/abs/2019MNRAS.488.4816W} {488, 4816}

\bibitem[\protect\citeauthoryear{{Wright}}{{Wright}}{2006}]{wright06}
{Wright} E.~L.,  2006, \mn@doi [\pasp] {10.1086/510102}, \href
  {https://ui.adsabs.harvard.edu/abs/2006PASP..118.1711W} {118, 1711}

\bibitem[\protect\citeauthoryear{{Wright} et~al.,}{{Wright}
  et~al.}{2010}]{wright10}
{Wright} E.~L.,  et~al., 2010, \mn@doi [\aj] {10.1088/0004-6256/140/6/1868},
  \href {http://adsabs.harvard.edu/abs/2010AJ....140.1868W} {140, 1868}

\bibitem[\protect\citeauthoryear{{Wyrzykowski} et~al.,}{{Wyrzykowski}
  et~al.}{2017}]{wyrzykowski17}
{Wyrzykowski} {\L}.,  et~al., 2017, \mn@doi [\mnras] {10.1093/mnrasl/slw213},
  \href {http://adsabs.harvard.edu/abs/2017MNRAS.465L.114W} {465, L114}

\bibitem[\protect\citeauthoryear{{Yang}, {Wang}, {Ferland}, {Yuan}, {Zhou}  \&
  {Jiang}}{{Yang} et~al.}{2013}]{yang13}
{Yang} C.-W.,  {Wang} T.-G.,  {Ferland} G.,  {Yuan} W.,  {Zhou} H.-Y.,
  {Jiang} P.,  2013, \mn@doi [\apj] {10.1088/0004-637X/774/1/46}, \href
  {http://adsabs.harvard.edu/abs/2013ApJ...774...46Y} {774, 46}

\bibitem[\protect\citeauthoryear{{Yang} et~al.,}{{Yang} et~al.}{2022}]{yang22}
{Yang} G.,  et~al., 2022, \mn@doi [\apj] {10.3847/1538-4357/ac4971}, \href
  {https://ui.adsabs.harvard.edu/abs/2022ApJ...927..192Y} {927, 192}

\bibitem[\protect\citeauthoryear{{van Velzen} et~al.,}{{van Velzen}
  et~al.}{2011}]{vanvelzen11}
{van Velzen} S.,  et~al., 2011, \mn@doi [\apj] {10.1088/0004-637X/741/2/73},
  \href {https://ui.adsabs.harvard.edu/abs/2011ApJ...741...73V} {741, 73}

\bibitem[\protect\citeauthoryear{{van Velzen}, {Mendez}, {Krolik}  \&
  {Gorjian}}{{van Velzen} et~al.}{2016}]{vanvelzen16b}
{van Velzen} S.,  {Mendez} A.~J.,  {Krolik} J.~H.,   {Gorjian} V.,  2016,
  \mn@doi [\apj] {10.3847/0004-637X/829/1/19}, \href
  {https://ui.adsabs.harvard.edu/abs/2016ApJ...829...19V} {829, 19}

\bibitem[\protect\citeauthoryear{{van Velzen}, {Pasham}, {Komossa}, {Yan}  \&
  {Kara}}{{van Velzen} et~al.}{2021a}]{vanvelzen21b}
{van Velzen} S.,  {Pasham} D.~R.,  {Komossa} S.,  {Yan} L.,   {Kara} E.~A.,
  2021a, \mn@doi [\ssr] {10.1007/s11214-021-00835-6}, \href
  {https://ui.adsabs.harvard.edu/abs/2021SSRv..217...63V} {217, 63}

\bibitem[\protect\citeauthoryear{{van Velzen} et~al.,}{{van Velzen}
  et~al.}{2021b}]{vanvelzen21}
{van Velzen} S.,  et~al., 2021b, \mn@doi [\apj] {10.3847/1538-4357/abc258},
  \href {https://ui.adsabs.harvard.edu/abs/2021ApJ...908....4V} {908, 4}

\makeatother
\end{thebibliography}
\bibliographystyle{mnras}


\label{lastpage}
\end{document}